\documentclass[12pt]{article}
\usepackage[margin=1in]{geometry}
\usepackage{amsmath,amsfonts,amssymb}
\usepackage{amsthm}
\usepackage{graphicx}
\usepackage{hyperref}
\usepackage[shortlabels]{enumitem}
\usepackage{fancyhdr}
\usepackage{titlesec}
\usepackage{abstract}
\usepackage{mathtools}
\usepackage{float}
\usepackage{tikz}
\usepackage{tikz-cd}

\usepackage{authblk}
\usepackage{amsthm}
\usepackage{setspace}
\usepackage{xcolor}
\usepackage{mathrsfs}
\usepackage{stmaryrd}
\usepackage{mathpartir}
\usepackage{dsfont}
\usepackage{listings}
\titleformat{\section}{\normalfont\Large\bfseries}{\thesection}{1em}{}
\titleformat{\subsection}{\normalfont\large\bfseries}{\thesubsection}{1em}{}

\newtheorem{definition}{Definition}[section]
\newtheorem{theorem}[definition]{Theorem}

\newtheorem{axiom}[definition]{Axiom}
\newtheorem*{theorem*}{Theorem}

\title{\textbf{Resolving CAP Through Automata-Theoretic Economic Design: A Unified Mathematical Framework for Real-Time Partition-Tolerant Systems}}

\author{\Large Dr. Craig S. Wright \\
\small Department of Computer Science, University of Exeter \\
\small \texttt{cw881@exeter.ac.uk}}

\date{}

\begin{document}

\maketitle

\begin{abstract}
\noindent The classical CAP theorem frames a foundational trade-off in distributed computing: a system cannot simultaneously guarantee consistency, availability, and partition tolerance. This paper reconceptualises the CAP trilemma through the lens of formal automata theory and strategic game-theoretic control. By modelling distributed systems as partition-sensitive deterministic automata, we expose the internal mechanics by which communication failures induce divergent state evolution. We extend this formalism with an economic layer wherein nodes behave as rational agents, governed by incentive-compatible mechanisms designed to enforce convergent behaviour under partial information and adversarial partitions. Our framework replaces binary CAP constraints with a quantifiable optimisation envelope, introducing the notion of $(\epsilon_C, \epsilon_A)$-bounded conformance. Consistency and availability are no longer treated as absolutes but as measurable properties within provable tolerance thresholds. Through constraint-preserving transitions, partition-aware arbitration functions, and penalisation of state divergence, we establish conditions under which resilient equilibria emerge and global correctness is preserved within bounded latency and deviation margins. We provide formal guarantees on liveness, convergence, and auditability, grounded in compositional bisimulation and category-theoretic semantics. In doing so, we transcend the traditional CAP boundary by embedding accountability and reversibility directly into the automaton structure and the strategic logic of its actors.
\end{abstract}

\vspace{1em}

\newpage
\tableofcontents
\newpage

\section{Introduction}

\subsection{Background and Motivation}

The CAP theorem, originally formulated by Brewer and later formalised by Gilbert and Lynch, establishes an impossibility result at the heart of distributed systems: that under network partition, a system must choose between consistency and availability. While this binary formulation has shaped the design of distributed protocols for decades, its categorical rigidity limits its applicability in modern, economically incentivised networks, particularly those where agents act strategically and communication is inherently unreliable.

Distributed ledgers, blockchain networks, and peer-to-peer protocols increasingly operate in adversarial environments where partitions are expected and economic incentives shape system behaviour. In such contexts, agents do not passively follow protocols—they optimise their own utility functions under uncertainty. The classical CAP framing lacks the expressive power to model these strategic, partition-aware behaviours and offers no mechanism for bounding degradation or incentivising recovery.

This work emerges from the need to bridge formal systems theory with economic mechanism design. Rather than accept the absolutes imposed by the original CAP theorem, we introduce a framework in which consistency and availability are relaxed to measurable $\epsilon$-bounded properties, and where partition tolerance is not merely endured but formally negotiated through reversible transitions and strategic penalties. By embedding economic logic into automaton transitions and formalising resilience through game-theoretic equilibria, we enable new classes of distributed systems that can reason over partitions, optimise outcomes under constraints, and formally verify bounded correctness despite adversarial disruption.

This integration reorients the problem from one of mutually exclusive guarantees to one of optimally managed trade-offs, where consistency and availability are dynamically negotiated rather than statically sacrificed. It sets the stage for an automata-based, incentive-aligned model of distributed computation where protocol correctness and economic rationality co-evolve under formal constraints.

\subsection{The CAP Theorem and its Limitations}

The CAP theorem states that no distributed system can simultaneously guarantee consistency, availability, and partition tolerance. In the presence of network partitions—where communication links between subsets of nodes are disrupted—a system must choose to either (1) remain consistent, ensuring all nodes agree on the same view of the data, or (2) remain available, responding to requests without coordination guarantees. This trilemma, although mathematically rigorous in its original formalisation, leads to rigid dichotomies that obscure practical realities in dynamic, economically driven systems.

The theorem assumes a static, binary failure model in which partitions either exist or do not, and nodes either respond or fail silently. It abstracts away notions of partial observability, incentive misalignment, and asynchronous link degradation, all of which are prevalent in real-world networks. Furthermore, the CAP theorem presumes passive components: nodes execute protocol steps without strategic intent or economic motivation. This idealisation fails in modern systems where agents may deviate, collude, or selectively comply based on perceived utility.

Additionally, CAP offers no gradation of failure: once a partition occurs, any violation of availability or consistency constitutes total failure with respect to that property. Such brittle semantics make it unsuitable for systems that tolerate degraded but bounded correctness, especially those that employ redundancy, delayed consistency, or transactional compensation. This limitation motivates a reframing of the CAP constraints into continuous, quantifiable metrics that can be optimised rather than strictly satisfied.

Our work addresses these issues by modelling distributed systems as automata whose transitions are constrained by communication topology and augmented by economic control logic. In this extended framework, the classical binary impossibility of CAP becomes a constraint-optimisation problem, where trade-offs between consistency and availability are formalised through $\epsilon$-bounded degradations and corrected via reversible transitions and incentive-aligned recovery mechanisms.

\subsection{Research Objectives and Scope}

This research aims to reframe the classical CAP theorem through the lens of automata theory, constraint optimisation, and economic mechanism design. Rather than treating consistency, availability, and partition tolerance as mutually exclusive properties, we introduce a formal system in which their trade-offs are expressed via quantitative relaxations. The primary objective is to define a partition-aware automaton model that incorporates adversarial network conditions while allowing for provable bounds on correctness and liveness within tolerable error margins.

To that end, we pursue the following specific goals: (1) to construct a formal model of distributed systems as distributed state machines subject to dynamic partitions; (2) to define a bounded relaxation of consistency and availability using $\epsilon$-conformance parameters; (3) to embed game-theoretic strategies into the system's evolution via economic incentives, audit mechanisms, and escrow-based enforcement; and (4) to demonstrate how these mechanisms can restore convergence and consistency even in adversarial settings.

The scope of this work encompasses formal specification of system behaviour under partitioned communication topologies, the design of incentive-compatible protocols for consistency maintenance, and the analysis of equilibrium strategies that are resilient to communication disruption. While we abstract from specific implementation details or platform dependencies, we ground our model in the context of blockchain-like systems and distributed ledgers where nodes act strategically and correctness is economically enforced. Simulation results and analytical proofs demonstrate how constraint-optimised trade-offs can supersede the rigid impossibility framing of classical CAP.

\subsection{Structure of the Paper}

The paper is organised into six core sections and four appendices. Section~\ref{sec:automata-model} develops the formal automaton-based framework for modelling distributed systems under adversarial partitions, introducing the Distributed System Automaton (DSA) and the semantics of partition-aware transitions. Section~\ref{sec:cap-reframing} mathematically reframes the CAP theorem, replacing binary impossibility with an $\epsilon$-bounded constraint optimisation model, and defines provable conformance conditions under adversarial network partitions. Section~\ref{sec:game-theory-control} embeds a game-theoretic economic layer, detailing how strategic node behaviour and economic audit mechanisms drive equilibrium stability and incentivise consistent behaviour. Section~\ref{sec:reconciliation} introduces the reversible and categorical structure of state reconciliation, establishing recovery conditions and state convergence logic. Section~\ref{sec:case-study} applies the formal model to a real-world case—Bitcoin Script—showing how incentive-compatible state machines align with convergent automata semantics. Section~\ref{sec:simulation} provides simulation data and empirical evidence supporting the feasibility and robustness of the proposed model under variable partition regimes and strategic deviation.

Appendix~A contains formal proofs of conformance theorems. Appendix~B includes formal strategy tables used in economic evaluation. Appendix~C documents simulation and code artefacts. Appendix~D formalises Merkle commitment logic.

\section{Axiomatic System Foundation}
\label{sec:system-model}
\label{sec:axiomatic-foundation}

This section establishes the formal basis for our distributed system model through a rigorous axiomatic framework. We begin by articulating foundational assumptions regarding node structure, local state spaces, and communication properties. These system assumptions, grounded in the canonical theory of asynchronous message-passing systems \cite{lynch1996distributed,attiya2004distributed}, are specified with precise constraints reflecting real-world network unreliability, adversarial partitioning, and the absence of global synchronisation.

From these assumptions, we construct a sequence of formal axioms governing global state composition, node-level transition determinism, partition-aware communication feasibility, and real-time responsiveness under bounded latency. Our model is aligned with interleaving semantics \cite{hoare1978communicating}, capturing concurrency via atomic, globally ordered events $\sigma_t \in \Sigma$, while respecting local visibility constraints and communication limitations imposed by dynamic connectivity graphs $C_t$.

\subsection{Reliability Model}\label{sec:reliability-model}

Each axiom is formulated to support the derivation of provable guarantees in Section~\ref{sec:reliability-model}, where we define and analyse correctness, liveness, and availability properties under network faults and adversarial conditions. The partition-aware transition function and bounded-latency responsiveness guarantee are especially designed to bridge the theoretical gap between abstract automata models and practically deployable, real-time distributed infrastructures \cite{gilbert2002brewer,dwork1988consensus}.

By explicitly codifying both the state-space semantics and event-level execution constraints, this axiomatic system serves as the foundation upon which our subsequent distributed automaton, economic coordination layer, and formal proof machinery are built. As such, it functions as both a logical base and a constraint system ensuring that all higher-level constructions remain consistent with observable and admissible system behaviour \cite{lamport1978time,alpern1985defining}.

\subsection{System Assumptions}

Here we establishe the closed-world specification for the formal system model. These assumptions exhaustively define the operational environment, agent capabilities, communication model, and adversarial constraints. They are exhaustive and exclusive: no behaviours, resources, or guarantees exist outside those explicitly stated herein. This foundation is essential for the validity of all axioms, formal derivations, and proofs of system properties in subsequent sections.

\begin{itemize}[label=--]

  \item \textbf{Nodes: Finite, Deterministic, Independent Agents}  
  Let $N = \{n_1, n_2, \ldots, n_k\}$ denote a finite, non-empty set of system nodes, with $k \in \mathbb{N}_{>1}$. Each $n_i \in N$ is an internally deterministic transition agent, i.e., a state machine with no randomness, oracle access, or external non-determinism. Formally, each node executes a total transition function $\delta_i: Q_i \times \Sigma_i \to Q_i$ where $Q_i$ is the node's local state space and $\Sigma_i$ the set of events visible to $n_i$. Nodes operate independently; there is no implicit synchronisation, no shared memory, and no access to global state. Coordination may only emerge from explicitly defined message exchanges.

  \item \textbf{Event Alphabet and Event Classes}  
  Let $\Sigma$ denote the global event alphabet. We assume $\Sigma$ is finite unless otherwise stated. Events are partitioned as:
  \[
  \Sigma := \Sigma_{\text{local}} \sqcup \Sigma_{\text{comm}} \sqcup \Sigma_{\text{fail}}
  \]
  with:
  \begin{itemize}
    \item $\Sigma_{\text{local}}$: local operations, e.g., $\texttt{compute}(v), \texttt{write}(x)$, affecting only internal node state.
    \item $\Sigma_{\text{comm}} = \{ \texttt{send}(m, n_i, n_j), \texttt{recv}(m, n_i, n_j) \mid n_i, n_j \in N, m \in \mathcal{M} \}$, where $\mathcal{M}$ is the structured message space (payloads, proofs, headers, etc.). These represent asynchronous communication primitives.
    \item $\Sigma_{\text{fail}}$: externally induced faults, including node crashes ($\texttt{crash}(n_i)$), link failures ($\texttt{drop}(n_i, n_j)$), or restarts.
  \end{itemize}
  A system execution trace is a finite interleaving of events over time:
  \[
  H := (\sigma_0, \sigma_1, \ldots, \sigma_T) \in \Sigma^T
  \]
  where $\sigma_t \in \Sigma$ is the (unique) global event at discrete step $t \in \mathbb{N}$. This models a globally interleaved sequence of events as observed by an omniscient trace recorder. It does not imply real-time synchrony but provides a well-ordered semantic framework for formal reasoning.

  \item \textbf{Local State Spaces and Transition Scope}  
  Each node $n_i$ maintains a local state $q_i \in Q_i$, where $Q_i$ is finite unless otherwise explicitly extended. The global system state is the product:
  \[
  q^t := (q_1^t, q_2^t, \ldots, q_k^t) \in Q := \prod_{i=1}^k Q_i
  \]
  Each node $n_i$ has a transition function $\delta_i: Q_i \times \Sigma_i \to Q_i$, where $\Sigma_i$ is the projection of events relevant to $n_i$ (i.e., for which $n_i$ is the acting or receiving principal). The composite global transition function $\delta: Q \times \Sigma \to Q$ is defined as:
  \[
  \delta(q, \sigma) := q' \text{ where } q_i' = \begin{cases}
    \delta_i(q_i, \sigma), & \text{if } \sigma \in \Sigma_i \\
    q_i, & \text{otherwise}
  \end{cases}
  \]
  This formulation ensures events are localised: only the affected nodes modify state; uninvolved nodes remain unchanged.

  \item \textbf{Communication Model and Channel Properties}  
  The system uses authenticated, asynchronous message-passing channels. Let $C \subseteq N \times N$ denote the set of potential directed communication links (typically $C = N \times N$). The following properties are assumed:

  \begin{enumerate}[label=(\alph*)]
    \item \textbf{Authenticity:}  
    All messages include unforgeable digital signatures. If $n_j$ receives message $m$ purportedly from $n_i$, then $n_j$ can cryptographically verify its origin. No adversary can impersonate a node under standard cryptographic assumptions.

    \item \textbf{Integrity:}  
    Messages are cryptographically hashed. Tampered messages are detected and discarded. No undetectable corruption occurs.

    \item \textbf{Non-FIFO Channels:}  
    Message delivery order is not guaranteed. Messages between $n_i$ and $n_j$ may arrive out-of-order.

    \item \textbf{Asynchronous Delivery with Bounded Delay (When Connected):}  
    If $(n_i, n_j) \in C_t$ at time $t$, then any message sent from $n_i$ to $n_j$ at time $t$ is delivered within $\Delta$ time units:
    \[
    \exists d \in [0, \Delta] \text{ such that } m \in \texttt{recv}_{n_j}(t + d)
    \]
    If $(n_i, n_j) \notin C_t$, then delivery is not guaranteed. Messages may be dropped, delayed indefinitely, or delivered after reconnection without ordering constraints.

    \item \textbf{Local Clocks and Temporal Semantics:}  
    Each node $n_i$ maintains a strictly increasing local clock $T_i: \mathbb{N} \to \mathbb{R}_{\geq 0}$. There is no synchronised global clock. Where causal ordering or temporal reasoning is required, nodes use logical clocks:
    \begin{itemize}
      \item \textit{Lamport clocks} $L_i(t)$ for establishing happened-before relations.
      \item \textit{Vector clocks} $\vec{V}_i(t) \in \mathbb{N}^k$ for causal comparison with partial order.
    \end{itemize}
    Clock drift, skew, and non-determinism in local time progression are assumed arbitrary but monotonic.
  \end{enumerate}

  \item \textbf{Partition Model and Adversary Constraints}  
  Let $P_t \subseteq C$ denote the partition function at time $t$, representing all links severed at that time. The active communication graph is:
  \[
  C_t := C \setminus P_t
  \]
  The connected components of $C_t$ define the live partitions:
  \[
  \mathscr{N}_t := \{ N_1^t, N_2^t, \ldots, N_m^t \} \text{ such that } \bigcup N_i^t = N, \quad N_i^t \cap N_j^t = \emptyset \text{ for } i \ne j
  \]
  Partitions may change arbitrarily over time, but must comply with a constraint model:
  \[
  \mathcal{P}_{\text{adv}} := \left\{ (P_t)_{t \in \mathbb{N}} \mid \Phi(P_t) \text{ holds } \forall t \right\}
  \]
  where $\Phi(P_t)$ encodes the allowed partition behaviours. Possible constraints include:
  \begin{itemize}
    \item \textbf{Bounded Link Disruption:} $|P_t| \leq B$ for some $B \in \mathbb{N}$.
    \item \textbf{Bounded Component Size:} $\forall i, |N_i^t| \geq \kappa$ for resilience threshold $\kappa$ (e.g., $\kappa > \lceil k/2 \rceil$).
    \item \textbf{Bounded Partition Duration:} $\forall t_0, \exists t_1 > t_0$ such that $P_{t_1} = \emptyset$ and $t_1 - t_0 \leq D_{\max}$.
    \item \textbf{Bounded Partition Frequency:} Over any interval $[t, t + T]$, the number of non-empty $P_s$ is at most $\lambda_{\max}$.
  \end{itemize}
  The adversary is modeled as a function:
  \[
  \mathcal{A}: \mathbb{N} \to 2^C \quad \text{such that } P_t = \mathcal{A}(t), \quad \mathcal{A} \in \mathcal{P}_{\text{adv}}
  \]
  The adversary is omniscient, network-aware, and reactive but cannot break cryptographic primitives, forge messages, or violate $\Phi$.
  
\end{itemize}

\subsection{Strategic Interaction}
\label{sec:strategic-interaction}

We now formalise the interface between distributed computation and game-theoretic decision-making. Each node $n_i \in N$ is modelled as a rational agent in a dynamic game played over the execution trace of the system. The game is played under \emph{incomplete information}—each node possesses a private type $\theta_i \in \Theta_i$—and \emph{restricted observability}, induced by adversarial partitions and asynchronous message arrival. These conditions introduce not only informational asymmetries but also strategic incentive misalignment. The structure of the game is therefore constrained not only by utility functions but also by the feasibility of state transitions within the Distributed System Automaton (DSA).

\subsubsection{Axiomatic Constraints on Play}

The axioms reiterated below establish the admissible domain of play. Each event $\sigma_t \in \Sigma$ must be partition-feasible and execute within bounded local response time. These constraints define the physical limitations of action and observation:

\begin{axiom}[Partition Impact]
\label{axiom:partition-impact}
Let $C \subseteq N \times N$ denote the set of all communication links, and $P_t \subseteq C$ the adversarially controlled partition at time $t$. The effective communication topology is:
\[
C_t := C \setminus P_t
\]
Define $\operatorname{Comm}(\sigma_t) \subseteq C$ as the set of links required by event $\sigma_t$ (non-empty only for $\sigma_t \in \Sigma_{\text{comm}}$). Then $\sigma_t$ is \emph{partition-feasible} at time $t$ iff:
\[
\operatorname{Comm}(\sigma_t) \subseteq C_t
\]
If this fails, then $\delta(q^t, \sigma_t)$ is undefined \cite{gilbert2002brewer}.
\end{axiom}

\begin{axiom}[Real-Time Constraint]
\label{axiom:real-time}
There exists a real-time bound $\tau > 0$ such that for any $n_i \in N$ and event $\sigma \in \Sigma_i$ at local time $T_i(t)$:
\begin{itemize}
  \item[(i)] $\sigma$ is rejected immediately if infeasible; or
  \item[(ii)] $\sigma$ completes execution on $n_i$ by $T_i(t) + \tau$.
\end{itemize}
For remote events, if $\operatorname{Comm}(\sigma) \ni (n_i, n_j)$, it is required that $(n_i, n_j) \in C_t$ and delivery respects latency $\Delta$ \cite{dwork1988consensus}.
\end{axiom}

\subsubsection{Game-Theoretic Integration}

Let the strategic interaction among nodes be modelled as a dynamic Bayesian game:
\[
\mathcal{G} := \left(N, \Sigma, A, T, \mathcal{F}, U, \mu\right)
\]
with the following elements:

\paragraph{Players}
$N = \{n_1, \dots, n_k\}$ is the finite set of rational agents (nodes), each seeking to maximise its expected utility.

\paragraph{Types}
Each $n_i$ has a private type $\theta_i \in \Theta_i$, drawn from a common prior $p(\theta) = \prod_i p_i(\theta_i)$. Types capture internal constraints, priorities, or honesty parameters.

\paragraph{Action Spaces}
Let $\Sigma_i \subseteq \Sigma$ be the set of events $n_i$ can initiate. Given the current system state $q^t$ and partition $P_t$, we define the \emph{admissible action set}:
\[
\Sigma_t^{\text{ad}} := \left\{ \sigma \in \Sigma \mid 
\begin{array}{l}
\sigma \in \Sigma_i \text{ for some } n_i \in N~\text{(C1)} \\
\operatorname{Comm}(\sigma) \subseteq C_t~\text{(C2)} \\
\text{execution latency}(\sigma) \le \tau~\text{(C3)}
\end{array}
\right\}
\]
Then for each player $n_i$, its feasible action space at time $t$ is:
\[
A_i^t := \left\{ \sigma \in \Sigma_i \mid \sigma \in \Sigma_t^{\text{ad}} \text{ and } \sigma \text{ consistent with } q_i^t \right\}
\]

\subsubsection{Information Filtration and Local Observability}

Each node $n_i$ has an observation history up to time $t$:
\[
\mathcal{F}_i^t := \sigma\left( \{ \sigma_0, \sigma_1, ..., \sigma_{t-1} \} \cap (\Sigma_i \cup \text{recv}_i) \right)
\]
This filtration $\mathcal{F}_i^t$ defines the player’s information set and hence the domain of its strategy function. Due to adversarial partitions and asynchrony, $\mathcal{F}_i^t$ need not contain globally consistent state.

\subsubsection{Strategies and the Interleaved Transition Mechanism}

Each agent selects a strategy $\pi_i$ which is a measurable function from its information set:
\[
\pi_i: \mathcal{F}_i^t \to A_i^t
\]
At each $t$, the profile $\pi = (\pi_1, ..., \pi_k)$ induces a tuple of proposed events $a^t = (a_1^t, ..., a_k^t)$. A global arbitration mechanism $\mu_t$ selects one atomic event to apply:
\[
\mu_t : A^t \to \Sigma_t^{\text{ad}} \cup \{\text{idle}\}
\]
If no $a_i^t$ are admissible, the idle transition is enacted. Otherwise, a single $\sigma_t \in \Sigma_t^{\text{ad}}$ is applied, producing:
\[
q^{t+1} = \delta(q^t, \sigma_t)
\]
The interleaving constraint renders the system a sequential extensive-form game under constrained concurrency.

\subsubsection{Payoffs and Equilibrium}

Each node $n_i$ possesses a belief-dependent utility function:
\[
U_i: \mathcal{H} \times \Theta \to \mathbb{R}
\]
mapping system histories $\mathcal{H}$ and type profiles $\theta$ to payoffs. Utilities may encode latency sensitivity, state consistency, or protocol adherence.

An equilibrium profile $\pi^* = (\pi_1^*, ..., \pi_k^*)$ satisfies the Bayesian Nash condition:
\[
\mathbb{E}_{\pi^*, \theta_{-i}} \left[ U_i(\mathcal{H}, \theta) \right] \ge 
\mathbb{E}_{\pi_i', \pi_{-i}^*, \theta_{-i}} \left[ U_i(\mathcal{H}', \theta) \right]
\]
for all $\pi_i' \ne \pi_i^*$ and for all $\theta_i \in \Theta_i$, where expectations are over the type prior and randomness in $\mu_t$ and message arrival.

\paragraph{Feasibility Constraint}
Only strategies mapping into physically realisable transitions are valid:
\[
\forall i, t:~ \pi_i(\mathcal{F}_i^t) \in \Sigma_t^{\text{ad}}
\]
This \emph{automaton-constrained admissibility} condition integrates game theory within the DSA execution semantics.

\paragraph{Outlook}
The strategic model introduced here establishes a foundation for incentive-compatible protocol design, detailed in Sections~\ref{sec:incentives} and \ref{sec:mechanism-design}, where we derive mechanism-theoretic constraints that align rational strategies with system-wide correctness guarantees.

\subsubsection{Repeated Interaction and Long-Run Incentives}
\label{sec:repeated-strategic-interaction}

We now extend the axiomatic interface to accommodate repeated strategic play over an unbounded time horizon. Each node $n_i \in N$ engages in an infinitely repeated game $\mathcal{G}_\infty$ where system dynamics, communication partitions, and payoff-relevant actions evolve over time. Let $t \in \mathbb{N}$ index discrete rounds, where each round corresponds to one automaton step with a valid transition $\delta(q^t, \sigma_t)$. A profile of local decisions, executed within bounded latency $\tau$ and constrained by communication feasibility, defines the admissible interaction set for every node.
\begin{axiom}[Real-Time Responsiveness]
\label{axiom:real-time-2}
For every node $n_i \in N$ and every admissible event $\sigma_t \in \Sigma$, there exists a bounded latency $\tau$ such that if $n_i$ is causally enabled to respond at time $t$, then its response action $a_i^{t'}$ is observable within $t' \leq t + \tau$. This axiom ensures that the system provides real-time responsiveness under bounded network and processing delays.
\end{axiom}

\begin{axiom}[Causal Continuity]
\label{axiom:causal-continuity}
Let $a_i^t \in A_i^t$ be the action taken by node $n_i$ at time $t$, and $q^t$ be the current system state. Then the evolution of $q^{t+1}$ satisfies:
\[
q^{t+1} = \delta(q^t, \mu_t(a^t)) \quad \text{iff } \mu_t(a^t) \in \Sigma_t^{\text{ad}}
\]
If $\mu_t(a^t) \notin \Sigma_t^{\text{ad}}$, then the automaton remains quiescent, i.e., $q^{t+1} = q^t$.
\end{axiom}

\begin{axiom}[Interaction Memory]
\label{axiom:interaction-memory}
Each node maintains a filtered event trace $\mathcal{F}_i^t$ constructed inductively as:
\[
\mathcal{F}_i^{t+1} := \mathcal{F}_i^t \cup \left\{ \sigma_t \mid \sigma_t \in \Sigma_i \cup \Sigma_{\text{recv}_i},~\text{deliverable under } C_t \right\}
\]
where $\Sigma_{\text{recv}_i} = \{ \texttt{recv}(m, n_j, n_i) \mid n_j \in N, m \in \mathcal{M} \}$. This defines the local information filtration under imperfect private monitoring. Partition-induced omission implies that $\mathcal{F}_i^t$ may exclude valid global events.
\end{axiom}

\begin{axiom}[Strategy Measurability]
\label{axiom:strategy-measurability}
Each node selects a strategy $\pi_i$ as a measurable mapping:
\[
\pi_i : \mathcal{F}_i^t \to A_i^t \subseteq \Sigma_i
\]
with the constraint that:
\[
\pi_i(\mathcal{F}_i^t) \in \Sigma_t^{\text{ad}}
\]
Nodes are assumed computationally capable of filtering actions for admissibility under the constraints defined by $\Sigma_t^{\text{ad}}$.
\end{axiom}

\begin{axiom}[Dynamic Consistency]
\label{axiom:dynamic-consistency}
Let $\delta \in (0,1)$ be a common discount factor. The long-run utility of node $n_i$ under profile $\pi = (\pi_1, ..., \pi_k)$ is:
\[
U_i(\pi) := \mathbb{E}_{\mu, \theta} \left[ \sum_{t=0}^{\infty} \delta^t u_i(q^t, \sigma_t, \theta) \right]
\]
where $u_i$ is the stage utility and expectation is over the type profile $\theta \sim p(\theta)$ and random arbitration. Utilities may encode delay aversion, finality, or protocol compliance.
\end{axiom}

\begin{axiom}[Equilibrium Compliance]
\label{axiom:equilibrium-compliance}
A strategy profile $\pi^* = (\pi_1^*, ..., \pi_k^*)$ is an automaton-constrained Bayesian Nash equilibrium if for all $n_i \in N$ and any alternative measurable strategy $\pi_i'$,
\[
U_i(\pi^*) \ge U_i(\pi_i', \pi_{-i}^*) \quad \text{subject to } \pi_i'(\mathcal{F}_i^t) \in \Sigma_t^{\text{ad}}~\forall t
\]
Only deviations consistent with axioms \ref{axiom:strategy-measurability} and \ref{axiom:real-time} are considered.
\end{axiom}

\paragraph{Folk Theorem Under Private Monitoring}
In the setting of imperfect private monitoring and sufficient patience ($\delta \to 1$), a broad class of individually rational payoff vectors above the minmax frontier can be sustained as sequential equilibria. This holds if players' filtrations are sufficiently informative to detect deviations through statistical inference. Conditions such as private differentiability and pairwise identifiability ensure the feasibility of punishment strategies within admissible transitions.

\begin{axiom}[Localised Reputation Enforcement]
\label{axiom:reputation}
Let $r_{ji}^t$ be the local reputation assigned to node $n_i$ by node $n_j$ at time $t$, with update function:
\[
r_{ji}^{t+1} := \Psi(r_{ji}^t, y_j^t(n_i), \mathcal{F}_j^t)
\]
Strategies may condition cooperation on these local beliefs:
\[
\pi_j(\mathcal{F}_j^t, r_{ji}^t) = \texttt{refuse}(n_i) \quad \text{if } r_{ji}^t < \rho
\]
where $\texttt{refuse}(n_i) \in \Sigma_j$ denotes admissible isolation behaviour.
\end{axiom}

\paragraph{Outlook}
This extension formalises long-run incentive structures in automaton-constrained systems under bounded observability and rationality. Mechanism design in subsequent sections will build on these axioms to construct enforceable cooperation protocols via locally observable punishment, repeated-game equilibria, and computationally tractable strategy profiles.\label{sec:incentives}\label{sec:mechanism-design}

\subsection{Axioms of State, Message, and Transition}

The semantics are defined over a closed-world model wherein every global system state and observable transition is externally visible and formally defined. Each step in the execution trace corresponds to a discrete, atomic event applied to a global state tuple, forming a total ordering over system evolution \cite{lynch1996distributed}. Although nodes maintain unsynchronised local clocks, the formal model adopts a discrete logical time $t \in \mathbb{N}$ to define global system dynamics \cite{lamport1978time}. These axioms serve as foundational primitives for the Distributed System Automaton in Section~\ref{sec:automata-model} and the game-theoretic construction in Section~\ref{sec:strategic-interaction}.

\begin{axiom}[State Composition]
\label{axiom:state-composition}
At each logical time $t \in \mathbb{N}$, the global system state is defined as:
\[
q^t = (q_1^t, q_2^t, \ldots, q_k^t) \in Q := \prod_{i=1}^k Q_i
\]
where $q_i^t \in Q_i$ is the complete local state of node $n_i$ at time $t$. The global state space $Q$ is the Cartesian product of the local state spaces. The system model is closed: all system-level dynamics are fully characterised by the tuple $q^t$ and the externally visible event $\sigma_t \in \Sigma$ applied at time $t$ \cite{hoare1978communicating}. No hidden or exogenous state variables exist outside this tuple.

\textbf{Implication:} This axiom enforces formal closure under composition. All analysis, trace semantics, and behavioural properties are derivable from $(q^0, \sigma_0, \sigma_1, \ldots)$ alone.
\end{axiom}

\begin{axiom}[Transition Determinism]
\label{axiom:transition-determinism}
Each node $n_i \in N$ defines a local deterministic transition function:
\[
\delta_i : Q_i \times \Sigma_i \to Q_i
\]
where $\Sigma_i \subseteq \Sigma$ denotes the subset of atomic events affecting $n_i$. The global transition function $\delta : Q \times \Sigma \to Q$ is defined compositionally as:
\[
\delta(q^t, \sigma_t) := q^{t+1} \quad \text{with} \quad q_i^{t+1} = \begin{cases}
  \delta_i(q_i^t, \sigma_t) & \text{if } \sigma_t \in \Sigma_i \\
  q_i^t & \text{otherwise}
\end{cases}
\]
for all $i \in \{1, \ldots, k\}$. The system enforces strict interleaving semantics: each $\sigma_t$ is a \emph{single atomic event} globally enacted at logical time $t$ \cite{alpern1985defining}. This reflects an event-driven, interleaved execution model where concurrency is represented as non-deterministic ordering of causally independent atomic steps \cite{attiya2004distributed}.

\textbf{Implication:} At each $t$, at most the nodes $n_i$ with $\sigma_t \in \Sigma_i$ may undergo local state change. All other $q_j$ remain unaffected. The system evolution is thus a deterministic function of the trace and the initial state. If future generalisation to simultaneous compound events is required, the alphabet $\Sigma$ may be lifted to a multiset model over $2^{\Sigma'}$.
\end{axiom}

\section{Formal Automaton Modelling of Distributed Systems}
\label{sec:automata-model}

To systematically capture the dynamics of a distributed system operating under uncertainty, network asynchrony, and adversarial conditions, we introduce a formal automaton-based framework. This model serves as the semantic foundation for analysing state transitions, message delivery, and strategic interaction among nodes within a partitioned communication topology. Each system component—nodes, channels, actions, and memory—is represented explicitly through a structured automaton, enabling the rigorous specification of transition semantics and admissibility conditions under both ideal and non-ideal execution environments.

The construction is layered, beginning with the local state space of each individual node and extending to global system dynamics mediated by a partition-aware transition function. The use of a partial transition function \( \delta_P \) allows us to constrain evolution to physically and causally valid trajectories, enforcing consistency with adversarial partitions and real-time delivery constraints. By distinguishing between node-level proposals and globally admissible events, we ensure the model accounts for the inherent informational asymmetry and coordination challenges in distributed systems.

This section formalises the syntax and semantics of the distributed system automaton (DSA), defines the structure and impact of communication partitions, and introduces the core mechanisms by which local actions are reconciled with global admissibility. These constructs collectively support the development of strategic models in later sections, enabling game-theoretic reasoning about system resilience, equilibrium behaviour, and adversarial robustness.

\subsection{Distributed System Automaton (DSA)}
\label{sec:dsa}

We now formalise the operational semantics underlying trace evolution under adversarial partitions and local rational strategies. The Distributed System Automaton (DSA) captures the global transition dynamics of the system and provides a foundation for admissibility, interleaving, and strategic interaction.

\begin{definition}[Distributed System Automaton]
\label{def:dsa}
A Distributed System Automaton is a tuple:
\[
\mathcal{A}_{DS} = (Q, \Sigma, \delta, q_0, N, C, (P_t)_{t \in \mathbb{N}})
\]
with components defined as follows:
\begin{itemize}
  \item $Q$ is the set of global system states.
  
  \item $\Sigma$ is the finite set of globally defined atomic events, partitioned into:
  \[
  \Sigma = \Sigma_{\text{local}} \cup \Sigma_{\text{comm}} \cup \Sigma_{\text{fault}}
  \]
  where:
  \begin{itemize}
    \item $\Sigma_{\text{local}} = \{ \texttt{comp}(n_i, \varphi) \mid n_i \in N,\ \varphi \in \Phi_i \}$ denotes local computation events; $\Phi_i$ is the set of local computation types for $n_i$.
    
    \item $\Sigma_{\text{comm}} = \{ \texttt{send}(m, n_i, n_j),\ \texttt{recv}(m, n_i, n_j) \mid (n_i, n_j) \in C,\ m \in \mathcal{M} \}$ denotes point-to-point communication events. The `recv` event is consistently represented as `$\texttt{recv}(m, \text{sender}, \text{receiver})$` throughout.
    
    \item $\Sigma_{\text{fault}} = \{ \texttt{crash}(n_i),\ \texttt{recover}(n_i) \mid n_i \in N \}$ denotes failure and recovery actions.
  \end{itemize}
  
  \item For each node $n_i \in N$, define its event projection $\Sigma_i \subseteq \Sigma$ as:
  \[
  \Sigma_i = \{ \sigma \in \Sigma \mid n_i\ \text{is the subject, sender, or receiver of}\ \sigma \}
  \]
  That is:
  \begin{itemize}
    \item $\texttt{comp}(n_i, \varphi) \in \Sigma_i$;
    \item $\texttt{send}(m, n_i, n_j) \in \Sigma_i$ and $\texttt{recv}(m, n_j, n_i) \in \Sigma_i$ for all $n_j \in N$;
    \item $\texttt{crash}(n_i),\ \texttt{recover}(n_i) \in \Sigma_i$.
  \end{itemize}

  \item $\delta: Q \times \Sigma \to Q$ is a partial deterministic transition function. Its domain is constrained to admissible event pairs $(q, \sigma)$ as defined by Axioms~\ref{axiom:partition-impact} and~\ref{axiom:real-time}.

  \item $q_0 \in Q$ is the initial system state.
  
  \item $N = \{n_1, \dots, n_k\}$ is the finite set of system nodes.
  
  \item $C \subseteq N \times N$ is the directed communication topology, where $(n_i, n_j) \in C$ permits transmission from $n_i$ to $n_j$.
  
  \item $(P_t)_{t \in \mathbb{N}}$ is the sequence of adversarial partition selections, where each $P_t \subseteq C$ represents the inactive links at time $t$, subject to constraints of the adversarial partition model $\mathcal{P}_{\text{adv}}$ (see Section~\ref{sec:system-model}).
\end{itemize}
\end{definition}

At each logical time step $t \in \mathbb{N}$, the automaton executes a single atomic event $\sigma_t \in \Sigma_t^{\text{ad}} \subseteq \Sigma$, where $\Sigma_t^{\text{ad}}$ denotes the admissible event set satisfying feasibility under $C_t = C \setminus P_t$ and respecting timing constraints (see Axioms~\ref{axiom:partition-impact} and~\ref{axiom:real-time}). The global transition at step $t$ is given by:
\[
q^{t+1} = \delta(q^t, \sigma_t)
\]
provided $\sigma_t$ is admissible. Events violating partition feasibility or exceeding the real-time bound are not included in $\Sigma_t^{\text{ad}}$ and are thus never selected.

Concurrency is captured by interleaving: while multiple events may be enabled at a given logical step, only one is selected nondeterministically or strategically by the arbitration mechanism defined in Section~\ref{sec:strategic-interaction}. This ensures determinism in the state evolution despite distributed autonomy. The DSA thus governs the admissible traces over which repeated-game reasoning and equilibrium enforcement are constructed.

\subsection{Local vs Global Transition Semantics}
\label{sec:local-vs-global-transitions}

Distributed systems operating under adversarial partitions and asymmetric information require a semantics that reconciles local actions proposed by nodes with the global constraints imposed by communication structure, causality, and state feasibility. In this section, we formalise the notion of a \emph{Partition-Aware Global Transition}, introduce the \emph{Admissible Event Set}, and define how locally generated proposals are filtered through a global arbitration function to determine valid state transitions. This framework ensures that system evolution adheres to the physical and logical bounds of the model.

\begin{definition}[Admissible Event Set]
\label{def:admissible-event-set}
Let $q \in Q$ be the global state at logical time $t$, and let $P_t \subseteq C$ denote the adversarial partition — the set of inoperative communication links at time $t$ — selected according to the adversarial model $\mathcal{P}_{\text{adv}}$. Define the effective communication graph $C_t := C \setminus P_t$. An event $\sigma \in \Sigma$ is \emph{admissible} under $(q, P_t)$, written $\sigma \in \Sigma^{\text{ad}}(q, P_t)$, if all of the following hold:
\begin{enumerate}[(i)]
  \item \textbf{Local Executability:} There exists a node $n_i \in N$ such that $\sigma \in \Sigma_i$, and the local state $q_i$ (the $i$-th component of $q$) satisfies the local preconditions of $\sigma$.
  \item \textbf{Communication Consistency:} If $\sigma$ involves communication, i.e., $\sigma \in \Sigma_{\text{comm}}$, then all edges in $\operatorname{Comm}(\sigma)$ must be active: $\operatorname{Comm}(\sigma) \subseteq C_t$.
  \item \textbf{Temporal Validity:} If $\sigma = \texttt{recv}(m, n_j, n_i)$, then there must exist a corresponding $\texttt{send}(m, n_j, n_i)$ at some prior $t' < t$, and the message $m$ must be in $n_i$'s input buffer at $t$ with $t - t' \leq \Delta$. Intermittent link failure during $[t', t]$ is permitted.
  \item \textbf{State Feasibility:} The global transition $\delta(q, \sigma)$ is defined, i.e., the result is a valid state in $Q$.
\end{enumerate}
\end{definition}

\begin{definition}[Partition-Aware Global Transition]
\label{def:partition-aware-transition}
The partition-aware transition function $\delta_P : Q \times \Sigma \times 2^C \rightharpoonup Q$ is defined as:
\[
\delta_P(q, \sigma, P_t) := 
\begin{cases}
  \delta(q, \sigma), & \text{if } \sigma \in \Sigma^{\text{ad}}(q, P_t), \\
  \text{undefined}, & \text{otherwise}.
\end{cases}
\]
This restricts the system's transition relation to those events that respect both physical constraints (partition $P_t$) and global admissibility. Since arbitration ensures only admissible events are enacted, the undefined branch is never taken during execution.
\end{definition}

\paragraph{Node-Level Proposal Semantics.}
Each node $n_i$ proposes a candidate event $a_i^t \in \Sigma_i$ based on its local state $q_i^t$ and memory $\mathcal{F}_i^t$. The set of locally feasible actions is:
\[
A_i^{\text{feasible}}(q_i^t, \mathcal{F}_i^t) := \left\{ \sigma \in \Sigma_i \ \middle|\ 
\begin{array}{l}
\sigma \text{ is executable from } q_i^t, \\
\operatorname{Comm}(\sigma) \text{ is consistent with } \mathcal{F}_i^t
\end{array}
\right\}.
\]
The full profile of proposed actions is $A^t := (a_1^t, \dots, a_k^t) \in \prod_{i=1}^k A_i^{\text{feasible}}(q_i^t, \mathcal{F}_i^t)$.

\paragraph{Arbitration.}
The arbitration function $\mu_t$ maps the joint proposal vector to an admissible global event:
\[
\mu_t : \prod_{i=1}^k A_i^{\text{feasible}}(q_i^t, \mathcal{F}_i^t) \to \Sigma^{\text{ad}}(q^t, P_t),
\]
with access to the full state $q^t$ and partition $P_t$. It ensures $\sigma_t := \mu_t(A^t)$ satisfies all conditions in $\Sigma^{\text{ad}}(q^t, P_t)$.

\paragraph{Enactment.}
The system advances according to:
\[
q^{t+1} := \delta_P(q^t, \sigma_t, P_t),
\]
with $\sigma_t$ guaranteed to be admissible.

\paragraph{Causal Coherence.}
Condition (iii) of admissibility ensures causal justification: any receive must be preceded by a corresponding send and arrive within $\Delta$. Transient failures are permitted; delivery suffices.

\paragraph{Semantic Summary.}
This layered model defines:
\begin{enumerate}[label=(\arabic*)]
  \item Local proposals from feasible local views.
  \item Arbitration selecting an admissible global event.
  \item Transition constrained by physical and logical semantics.
\end{enumerate}
The partial function $\delta_P$ enforces boundaries of lawful evolution. Only events consistent with adversarial conditions, causal dependencies, and state constraints are enacted.

\subsection{Partition Function Definition}
\label{sec:partition-function}

To rigorously formalise the impact of adversarial behaviour on the network’s topology, we define the partition function as the primary construct specifying which communication links are inactive at each logical time step.

\begin{definition}[Partition Function]
Let \( N \) be the set of nodes and \( C \subseteq N \times N \) the set of all potential communication links. The adversarial partition at time \( t \in \mathbb{N} \) is the set
\[
P_t \subseteq C,
\]
representing the communication links that are \emph{inactive} at time \( t \) due to failures or adversarial control. These are precisely the links through which no message can be reliably transmitted at that time step.
\end{definition}

\paragraph{Effective Topology.}
Given the adversarial partition \( P_t \), we define the effective communication topology at time \( t \) as
\[
C_t := C \setminus P_t,
\]
the set of all communication links that are active and operational at time \( t \). All feasible communication events must be constrained to edges in \( C_t \), ensuring that message-passing semantics respect the prevailing network constraints.

\paragraph{Temporal Partition Map.}
The sequence of adversarial partitions across time is defined as
\[
\mathcal{P}_{\text{adv}} := \left\{ (P_t)_{t \in \mathbb{N}} \ \middle|\ \forall t,\ P_t \subseteq C \text{ and } \Phi(P_t) \text{ holds} \right\},
\]
where \( \Phi(P_t) \) denotes a predicate encoding admissibility constraints on the adversary’s power—such as temporal continuity, topological invariants, or bounds on the maximum number of disrupted links. This formalism supports both static and dynamic adversarial models, providing a basis for verifying system behaviour under non-ideal and evolving network conditions.

\subsection{State Space Explosion and Composition}

A fundamental challenge in modelling distributed systems via automata lies in the phenomenon of state space explosion. Since the global state space \( Q \) is constructed as the Cartesian product of the local state spaces \( Q_1, Q_2, \ldots, Q_k \) of all \( k \) participating nodes, its cardinality grows exponentially:
\[
|Q| = \prod_{i=1}^k |Q_i|.
\]
This exponential blow-up renders direct computation or enumeration of the global transition system intractable for even modestly sized networks, especially when each node maintains a non-trivial internal state.

To mitigate this complexity, structured composition techniques are employed. Hierarchical automata allow for the abstraction and encapsulation of local subsystems, enabling modular reasoning about system components and reducing interleaving overhead in the composed transition graph. Moreover, compositional bisimulation provides a powerful formal tool for verifying behavioural equivalence between a full system and its reduced or abstracted representation. Through bisimulation-preserving composition, it is possible to substitute subsystems with simpler equivalents while preserving observable behaviour, thereby enabling scalable analysis without loss of semantic fidelity.

These reductions support compositional verification, where local correctness proofs are lifted to the global system by exploiting the preservation of temporal properties under bisimulation. Furthermore, automata-theoretic encodings of local strategy synthesis, such as symbolic fixpoint iteration or temporal abstraction refinement, are feasible only under aggressive state-space reduction. Without such mechanisms, the synthesis of admissible executions under adversarial partitions becomes computationally prohibitive.

To address this, we adopt a multi-layered modelling approach. At the lowest level, we retain full local-state fidelity per node, preserving fine-grained transition semantics. At intermediate layers, we factor behaviour into equivalence classes modulo local observational indistinguishability, yielding quotient automata per node. The global system is then reconstructed compositionally from these quotient components. This yields a balance between precision and tractability, enabling scalable symbolic execution, equilibrium selection, and verification across adversarially partitioned system trajectories.

In our framework, the choice of composition semantics—interleaved, synchronous, or hybrid—further influences explosion severity and the admissibility of local-to-global behavioural lifting. Our reductions ensure semantic closure under transition simulation and enable modular refinement under network topology constraints imposed by \( P_t \). This modularity is essential for real-world distributed systems subject to dynamic link failure, strategic interaction, and adversarial interference.

\section{Mathematical Reframing of the CAP Theorem}\label{sec:cap-reframing}

The CAP theorem, classically stated, asserts that no distributed system can simultaneously guarantee Consistency, Availability, and Partition Tolerance in the presence of network faults. This informal trilemma has historically driven the design of fault-tolerant systems by forcing a choice between consistency and availability under partition. However, its informal framing introduces ambiguity in both scope and operational interpretation, obscuring the quantitative nature of the trade-offs.

In this section, we reconstruct the CAP theorem as a formal optimisation problem over adversarial partitions. Rather than treating Consistency and Availability as binary, all-or-nothing properties, we define them as quantitatively measurable functions bounded by tolerances \((\theta_C, \theta_A)\). Partition Tolerance is reframed as a constraint set: a class of permitted link failures \(\mathcal{P}_{\text{adv}}\) that represents the adversary's action space.

This reformulation enables the derivation of feasibility regions in the system’s design space where degraded, yet provable, CAP conformance is possible. Through this lens, the CAP theorem is not a hard impossibility but a constrained optimisation problem: given an adversarial partition schedule, can we construct a system whose consistency deviation \(\epsilon_C\) and response delay \(\epsilon_A\) remain bounded?

We introduce the machinery to express consistency and availability as constraint functions over histories and executions, and then demonstrate that the feasible set of strategies \(\Pi\) and implementations \(I\) satisfying these constraints under all \(P_t \in \mathcal{P}_{\text{adv}}\) is non-empty for a wide class of protocols. This mathematical reframing thus lays the foundation for synthesising distributed systems that are both resilient and optimally performant within adversarially bounded environments.

\subsection{Formal Definitions}

\begin{definition}[Consistency (Linearizability)]
Let \( H \) be a finite history of operations over a shared object, consisting of invocation and matching response events. We say \( H \) is \emph{linearizable} if there exists a total order \( \prec \) on the completed operations in \( H \) such that:
\begin{enumerate}
    \item \textbf{Real-Time Precedence Preserved:} If operation \( op_1 \) completes before \( op_2 \) begins in real time (i.e., \( resp(op_1) < inv(op_2) \)), then \( op_1 \prec op_2 \).
    \item \textbf{Specification Conformance:} The total order \( \prec \) produces a legal sequential execution according to the object’s specification, where each operation appears to take effect atomically at some point between its invocation and response.
\end{enumerate}
This provides a strong form of consistency, ensuring the system behaves as if all operations were executed instantaneously in a sequential order respecting real-time constraints.
\end{definition}

\begin{definition}[Availability with Response Bound]
Let \( \Sigma \) denote the set of permitted operations and \( P_t \) the adversarial partition at logical time \( t \). A system is said to be \emph{available with response bound \( \tau \)} under \( P_t \) if:
\[
\forall op_i \in \Sigma, \quad \exists t_i \in \mathbb{N} \text{ such that } \text{resp}(op_i, P_t) - \text{inv}(op_i) \leq \tau
\]
That is, every operation \( op_i \) submitted to the system during partition \( P_t \) must receive a response within a bounded time \( \tau \), regardless of network degradation or link failure. Availability is guaranteed even under partial network failure, provided the conditions on response latency hold.
\end{definition}

\begin{definition}[Partition Tolerance]
A distributed system is said to exhibit \emph{partition tolerance} with respect to an adversarial partition model \( \mathcal{P}_{\text{adv}} \) if there exists a partition-aware transition function \( \delta_P : Q \times \Sigma \times \mathcal{P}_{\text{adv}} \rightharpoonup Q \) such that:
\[
\forall P_t \in \mathcal{P}_{\text{adv}}, \quad \exists \sigma_t \in \Sigma^{\text{ad}}(q^t, P_t) \text{ such that } \delta_P(q^t, \sigma_t, P_t) \text{ is defined}.
\]
This guarantees that for every allowable partition \( P_t \), the system can make progress—i.e., it remains live—by executing at least one admissible event \( \sigma_t \). The model ensures that no partition in \( \mathcal{P}_{\text{adv}} \) can completely stall system evolution.
\end{definition}

\subsection{Relaxed CAP via Constraint Optimisation}

To address the fundamental limitations of the classical CAP theorem, we adopt a quantitative relaxation framework in which consistency and availability are treated as continuous variables bounded by tolerances. This allows for precise trade-off analysis and adaptive design within adversarial or degraded environments.

\begin{definition}[$\epsilon$-CAP Conformance]
Let \( \theta_C \) and \( \theta_A \) be fixed system tolerances for consistency and availability, respectively. A distributed system is said to satisfy \emph{\((\epsilon_C, \epsilon_A)\)-CAP conformance} under partition \( P_t \) if the following hold:
\begin{align*}
\epsilon_C &= \max_{i,j} \left| \text{real}(op_i, op_j) - \text{linear}(op_i, op_j) \right| \leq \theta_C, \\
\epsilon_A &= \max_{i} \left( \text{resp}(op_i) - \tau \right) \leq \theta_A.
\end{align*}

\begin{itemize}
    \item \textbf{Consistency Deviation \( \epsilon_C \):} Measures the maximum deviation between the real-time precedence of operations and their corresponding order in a legal linearization. This quantifies staleness or reordering violations and generalises linearizability by bounding rather than eliminating inconsistency.
    \item \textbf{Availability Deviation \( \epsilon_A \):} Measures the maximal response delay over a defined response deadline \( \tau \), reflecting bounded liveness under partial synchrony or disruption.
\end{itemize}

This formulation enables system designers to define admissible operational envelopes where both consistency and availability are not treated as binary properties, but as optimisable constraints within quantifiable margins.
\end{definition}

\subsection{Feasibility Region Expansion}

A central objective in resilient distributed systems is to identify strategy–implementation pairs that provably satisfy relaxed CAP constraints under all admissible partitions. Rather than treating consistency and availability as antagonistic absolutes, we construct a conformance envelope wherein both properties are met within specified bounds across the full adversarial partition space.

\begin{theorem}[Provable Conformance Under Partition]
Let \( \mathcal{P}_{\text{adv}} \) be the admissible set of adversarial partitions, and let \( (\theta_C, \theta_A) \) be system-defined tolerances for consistency and availability deviations. Then there exists a pair \( (\Pi, I) \), where \( \Pi \) is a distributed strategy profile and \( I \) an implementation class, such that for all \( P_t \in \mathcal{P}_{\text{adv}} \),
\[
\mathcal{A}_{DS}^{\Pi, I} \models (\epsilon_C \leq \theta_C \land \epsilon_A \leq \theta_A).
\]
\begin{itemize}
    \item \( \mathcal{A}_{DS}^{\Pi, I} \) denotes the automaton induced by strategy \( \Pi \) and implementation \( I \) over the distributed system.
    \item The satisfaction relation \( \models \) indicates that every admissible execution trace under partition schedule \( (P_t) \) conforms to the bounded consistency–availability envelope.
\end{itemize}

This theorem defines a computable feasibility region within the strategy–implementation space. It establishes the existence of bounded-resilience designs that operate safely within adversarial settings and underpins constraint-optimised synthesis procedures explored in later sections.
\end{theorem}

\section{Economic Layer: Game Theoretic System Control}\label{sec:game-theory-control}

This section formalises the integration of economic incentives and game-theoretic constraints into the control layer of a distributed system under adversarial network partitions. Each node in the system is modelled as a rational agent participating in a strategic environment shaped by communication constraints, information asymmetry, and audit-enforced accountability.

The design objective is to align each node's strategic incentives with protocol adherence, even when communication is disrupted or when malicious coalitions form. By embedding utility-maximising behaviour directly into the transition logic, the system ensures that compliance is a rational outcome and deviation is financially or operationally disadvantageous.

Through the definitions of strategic interaction over partitions, partition-resilient equilibria, and incentive-compatible mechanisms involving audit, penalty, and escrow, we establish a formal architecture in which equilibrium behaviour reinforces consistency, availability, and verifiability within the distributed system. This economic layer transforms enforcement from centralised supervision to decentralised, provable, and self-interested compliance.

\subsection{Nodes as Strategic Agents}
\begin{definition}[Game over Partitions]
Let \( G = (N, S, U, P_t) \) define a game-theoretic model over an adversarial partitioned network, where:
\begin{itemize}
  \item \( N = \{n_1, \dots, n_k\} \) is the set of agents (nodes) in the distributed system;
  \item \( S = S_1 \times \dots \times S_k \) is the joint strategy space, with \( S_i \) denoting the local strategy set of agent \( n_i \);
  \item \( U_i : S \to \mathbb{R} \) is the utility function of agent \( n_i \), mapping joint strategies to real-valued outcomes based on local objectives (e.g., consistency reward, latency penalties);
  \item \( P_t \in \mathcal{P}_{\text{adv}} \) is the adversarial partition at logical time \( t \), restricting inter-agent communication.
\end{itemize}
\end{definition}

This model frames the system as a multi-agent game played under communication constraints induced by the adversary. Agents select local strategies based on their limited view and communication availability, and their payoffs depend not only on their own actions but also on the joint behaviour of others, modulated by the current partition \( P_t \). This allows for the formal analysis of incentive compatibility, equilibrium robustness, and optimal policy synthesis under partition-induced uncertainty.

\subsection{Utility Functions}

Each node \( n_i \in N \) is modelled as a rational agent maximising a utility function that evolves over time and responds to both local behaviour and global protocol constraints. The utility at logical time \( t \) is given by:

\[
U_i^t = R_i^t - C_i^t + P_i^t
\]

\begin{itemize}
  \item \( R_i^t \): The reward component, representing gains obtained through successful participation in globally consistent actions. This may include bonuses for contributing to linearizable histories, advancing consensus, or aiding in system availability despite partitions.
  
  \item \( C_i^t \): The cost component, capturing resource expenditure (e.g., computation, communication latency, message drops) and opportunity loss from aborted or delayed actions due to partition constraints.
  
  \item \( P_i^t \): The penalty term, imposed through an economic audit protocol that monitors and punishes strategic deviations, such as consistency violations, selfish behaviour, or protocol manipulation. This models externally enforced compliance (e.g., through cryptographic accountability or contract enforcement).
\end{itemize}

This formulation supports equilibrium analysis under adversarial partitions, allowing one to quantify the incentive alignment of consistency-preserving behaviour versus partition-exploiting deviation. It also enables formal guarantees over utility-bounded rational actors operating within distributed consensus protocols.

\subsection{Partition-Resilient Equilibria}

In partitioned distributed systems, rational agents must select strategies not only in light of others' actions, but also under the influence of dynamic communication failures. This motivates a refinement of classical Nash equilibrium that accounts for adversarially evolving partitions.

\begin{definition}[Resilient Nash Equilibrium]
A joint strategy profile \( S^* = (S_1^*, \ldots, S_n^*) \) is a \emph{resilient Nash equilibrium} under partition \( P_t \) if and only if:
\[
\forall i \in N, \quad U_i(S_i^*, S_{-i}^*, P_t) \geq U_i(s_i', S_{-i}^*, P_t), \quad \forall s_i' \in S_i.
\]
\end{definition}

This condition guarantees that, for any node \( n_i \), no unilateral deviation in strategy \( s_i' \) yields higher utility given the current partition \( P_t \) and the strategies of all other nodes. Resilience here captures the system’s strategic stability in the face of evolving communication failures, adversarial disruptions, or rational exploitation. It ensures that nodes have no incentive to defect even when the network is fragmented, and aligns utility-optimal behaviour with protocol-conformant operation.

Such equilibria serve as fixed points of rational behaviour under network uncertainty, offering a formal foundation for incentive-compatible protocol design in adversarial distributed systems.

\subsection{Mechanism Design: Audit, Penalty, Escrow}

To ensure strategic conformity in adversarially partitioned distributed systems, we introduce a mechanism \( g: S \to Q \) that maps strategy profiles to system-level outcomes. The mechanism is constructed to enforce incentive compatibility, discourage collusion, and ensure economic accountability.

\begin{itemize}
    \item \textbf{Truthful Reporting Incentive:} The mechanism \( g \) is designed such that each agent \( n_i \) maximises its expected utility by reporting its local state and observations truthfully:
    \[
    s_i^{\text{truth}} \in \arg\max_{s_i \in S_i} U_i(s_i, S_{-i}, P_t).
    \]

    \item \textbf{Collusion Penalty:} For any non-truthful collusive strategy subset \( S' \subset S \) involving multiple nodes, it holds that:
    \[
    \sum_{i \in N'} U_i(S') < \sum_{i \in N'} U_i(S^*),
    \]
    where \( S^* \) denotes the truthful equilibrium profile and \( N' \subseteq N \) the set of colluding agents.

    \item \textbf{Escrow-Based Economic Accountability:} Each agent \( n_i \) is required to post an escrow deposit \( e_i \). If a node is found to have propagated an incorrect or inconsistent state (as verified by an embedded economic audit protocol), then:
    \[
    e_i \rightarrow \text{forfeit}.
    \]
    The audit procedure detects protocol violations by checking consistency of the reported state trajectories against admissible event histories and observable transitions under \( P_t \). Escrow forfeiture is triggered upon provable deviation.

\end{itemize}

This mechanism guarantees that rational agents are disincentivised from deviation, incentivised towards full compliance, and financially penalised for misbehaviour. The economic structure embeds protocol-enforceable game-theoretic constraints into the system’s transition semantics.

\section{Automata-Based State Reconciliation}\label{sec:reconciliation}

In distributed systems subject to partitions, adversarial communication loss, and asynchronous execution, maintaining coherent global state becomes non-trivial. Divergence arises when nodes evolve independently under partial information, leading to inconsistencies in local views and potential violations of system-wide invariants. To address this, we formalise reconciliation as a structured recovery process within an automata-theoretic framework.

Automata-based reconciliation provides a mathematically rigorous mechanism for analysing and repairing state divergence. By identifying divergent configurations, defining recovery-specific transition subsets, and verifying convergence properties, the framework ensures that systems remain provably capable of recovering from inconsistency under defined adversarial conditions. Recovery transitions are treated as formal constructs—either as explicit recovery rules or as reversals of permissible forward transitions—ensuring compliance with application-level semantics and safety properties.

The following subsections define the structure of divergent states, the design of recovery automata, and the properties necessary to guarantee convergence. Reconciliation is modelled as a sequence of transitions over the state space \( Q \), constrained by a recovery map \( \delta_{\text{rev}} \) and terminating in a consistent state \( q' \in \mathscr{C} \). This model supports formal verification, incentive-compatible audits, and the synthesis of system-level resilience under bounded rationality and strategic behaviour.

\subsection{Reversible Automata Formalism}

\begin{definition}[Reversible Automaton]
An automaton \( \mathcal{R} = (Q, \Sigma, \delta, q_0) \) is said to be \emph{reversible} if for every reachable state \( q' \in Q \), there exists a state \( q \in Q \) and an input symbol \( \sigma \in \Sigma \) such that
\[
\delta(q, \sigma) = q'.
\]
\end{definition}

Reversibility formalises the condition that every state in the automaton can be reached from some prior state via a well-defined transition. This ensures that the system's evolution is not only forward-executable but also structurally invertible in terms of trace recovery, although not necessarily uniquely so.

In the context of distributed system automata, reversibility supports auditability and verifiability of state transitions under adversarial conditions. The model allows for post-facto reconstruction of execution paths, critical to enforcement mechanisms and the embedding of rollback-aware semantics into the economic layer. Combined with the escrow and penalty logic, reversible automata underpin provable accountability through traceable execution semantics even when communication links are disrupted or Byzantine behaviour is observed.

\subsection{Divergence State Definitions}

\begin{definition}[Divergent State Set]
Let \( Q \) be the global state space defined as the product of local states \( Q_1 \times \cdots \times Q_k \). The divergent state set at time \( t \) is defined as:
\[
\mathscr{D}_t = \left\{ q \in Q \,\middle|\, \exists i, j \in \{1, \ldots, k\} : q_i \neq q_j \land \neg\operatorname{Sync}(q_i, q_j) \right\},
\]
where \( \operatorname{Sync}(q_i, q_j) \) holds if the local timestamps and transition histories of \( q_i \) and \( q_j \) are mutually consistent within protocol-defined bounds.
\end{definition}

Divergence captures the semantic deviation of local views within a distributed system due to communication failures, delays, or adversarial disruption. The condition \( q_i \neq q_j \) is necessary but not sufficient; divergence is marked only when observable disagreement violates expected synchronisation—typically modelled by timestamp misalignment, version skew, or committed history mismatch.

Tracking \( \mathscr{D}_t \) is critical for conflict resolution, state repair, and rollback initiation in systems with reversibility and economic enforcement layers. It forms the basis for defining convergence guarantees, bounding eventual consistency, and penalising delayed reconciliation or malicious divergence propagation.

\subsection{Recovery Protocol}

\begin{definition}[Recovery Transition Subset]
Let \( \delta_{\text{rev}} \subseteq \delta \) be the set of designated recovery transitions. These transitions are explicitly constructed to resolve divergence by driving the system from any inconsistent global state to a consistent one. Formally, the protocol satisfies:
\[
\forall q \in \mathscr{D}_t,\, \exists q' \in Q : \delta_{\text{rev}}^*(q) = q' \land q' \in \mathscr{C},
\]
where \( \mathscr{D}_t \subset Q \) is the divergent state set at time \( t \), \( \mathscr{C} \subset Q \) is the set of converged states, and \( \delta_{\text{rev}}^* \) denotes the finite closure of recovery transitions applied to \( q \).
\end{definition}

The recovery protocol ensures that despite transient partitions or adversarial disruptions, any reachable divergent state can be transitioned into a globally consistent configuration through a finite sequence of recovery steps. These transitions may include reconciliatory message exchanges, ledger replays, compensating writes, or coordinated rollbacks.

Design of \( \delta_{\text{rev}} \) must satisfy both soundness (no false convergence) and completeness (all divergences are resolvable), forming the operational backbone for reversibility, audit compliance, and post-partition consistency guarantees in strategic distributed systems.

\section{Category-Theoretic Structure of Consistency}

The behaviour of distributed systems under communication constraints admits a natural categorical formulation, where system states, transitions, and partition effects are represented as structured morphisms and functors. Let \( \mathcal{C}_{DS} \) denote a base category whose objects are global system states and whose morphisms are transition-inducing operations—either local or composed from multiple concurrent local actions.

Consistency constraints such as linearizability and sequential consistency manifest as diagrammatic properties. In particular, linearizability corresponds to the requirement that all transition diagrams commute—i.e., the outcome of a sequence of operations is invariant under permissible reorderings that respect real-time precedence:
\[
f \circ g = g \circ f \quad \Rightarrow \quad \text{Converged state independent of operational interleaving}.
\]

Partitions are formally represented via endofunctors or quotient constructions over \( \mathcal{C}_{DS} \), producing a subcategory \( \mathcal{C}_{DS}^{\text{partitioned}} \). The inability to synchronise or observe operations across partition boundaries obstructs morphism composition. Functors \( F: \mathcal{C}_{DS}^{\text{partitioned}} \to \mathcal{C}_{DS} \) that attempt to lift execution fragments into globally linearised histories often fail to preserve limits or colimits, reflecting the impossibility of achieving certain consistency properties under adversarial partitions.

To reason about reconciliation, we study diagram completion in categories with partial morphisms or enriched with delay monads. Commutativity failures induce a search for mediating morphisms that restore equivalence. Recovery automata then define a cone or cocone over diverging paths, enabling resolution via colimit construction.

In this framework, consistency becomes a question of the existence and stability of commutative diagrams across indexed categories of state under varying communication topologies. The category-theoretic lens thus provides both a structural abstraction and a precise language for expressing and analysing consistency constraints in partitioned distributed systems.

\subsection{System States as Category Objects}

To enable algebraic reasoning over distributed system evolution, we introduce a categorical abstraction. Let \( \mathcal{C}_{DS} \) be a small category defined as follows:

\begin{itemize}
  \item \textbf{Objects:} Each object \( q \in \text{Ob}(\mathcal{C}_{DS}) \) corresponds to a global system state in the automaton \( \mathcal{A}_{DS} \).
  \item \textbf{Morphisms:} For any two objects \( q, q' \in \text{Ob}(\mathcal{C}_{DS}) \), a morphism \( f : q \to q' \) exists if there is a transition \( \delta(q, \sigma) = q' \) for some event \( \sigma \in \Sigma \). Thus, morphisms represent feasible state transitions induced by admissible events.
\end{itemize}

Composition of morphisms follows the transition closure of the automaton:
\[
f : q \to q', \quad g : q' \to q'' \quad \Rightarrow \quad g \circ f : q \to q''
\]
with identity morphisms \( \text{id}_q : q \to q \) corresponding to stutter steps (i.e., null transitions or time progression without action).

This categorical formulation allows us to represent distributed computations as composable morphisms over system states and to apply categorical constructions (e.g., functors, limits, monoidal structure) for structural and behavioural analysis. Such abstraction is crucial for defining equivalence classes of executions, applying homological invariants, and designing correctness-preserving transformations over adversarially evolving topologies.

\subsection{Consistency as Commutativity}

In the categorical model \( \mathcal{C}_{DS} \), consistency properties such as linearizability can be expressed through the algebraic structure of morphism composition.

\begin{definition}[Commutativity and Convergence]
Let \( f : q \to q' \) and \( g : q \to q'' \) be morphisms representing distinct operation sequences from a common state \( q \). If there exists a common successor state \( q^* \) such that both sequences composed in either order reach \( q^* \), i.e.,
\[
f \circ g = g \circ f : q \to q^*,
\]
then the system is said to exhibit operation-level commutativity at \( q \), and the corresponding transitions are order-independent.
\end{definition}

This commutativity implies that the final state of the system is invariant under the interleaving of concurrent operations, which is a necessary condition for linearizability. In particular, if for all concurrent operations such commutative diagrams exist and preserve the global specification, then the system maintains strong consistency guarantees.

Thus, consistency in this framework is captured by the existence of commuting diagrams over state morphisms, ensuring that independent execution schedules yield equivalent and converged outcomes.

\subsection{Partitions as Functorial Obstructions}

Let \( \mathcal{C}_{DS}^{\text{partitioned}} \) be the category of distributed system states and transitions under partition constraints, and \( \mathcal{C}_{DS}^{\text{global}} \) the corresponding category assuming full communication and synchrony. A functor
\[
F : \mathcal{C}_{DS}^{\text{partitioned}} \to \mathcal{C}_{DS}^{\text{global}}
\]
maps partition-constrained execution histories to their globally idealised counterparts, preserving the object structure (states) and, when possible, morphisms (transitions).

However, partition-induced loss of morphism composability—due to message omission, delay, or reordering—imposes functorial obstructions. In categorical terms, not all commutative diagrams in \( \mathcal{C}_{DS}^{\text{global}} \) lift to corresponding diagrams in \( \mathcal{C}_{DS}^{\text{partitioned}} \). That is, the functor \( F \) fails to preserve certain pullbacks or limits:
\[
\text{If } \begin{tikzcd}
& q' \arrow[dr, "g"] \\
q \arrow[ur, "f"] \arrow[rr, dashed, "g \circ f"'] && q''
\end{tikzcd}
\text{ commutes in } \mathcal{C}_{DS}^{\text{global}}, \text{ it need not in } \mathcal{C}_{DS}^{\text{partitioned}}.
\]

This breakdown captures the essential obstruction that partitions introduce: morphisms (operations) that would compose into a consistent, linearised global state may no longer be composable or even defined in the partitioned model. As a result, reconciliation protocols must either re-establish the missing morphisms or construct equivalence classes over non-commuting paths, inducing a form of eventual consistency via colimits.

Thus, analysing the lifting properties of \( F \) characterises the boundary between provable consistency and fundamental limitations imposed by adversarial network conditions.

\section{Case Study: Bitcoin Script as Convergent Automata}\label{sec:case-study}

Bitcoin Script provides a constrained, decidable execution layer designed for predictable and verifiable evaluation across distributed nodes. Despite its non-standard Turing-complete nature, the script system can be rigorously modelled as a convergent automaton: a formal machine whose execution deterministically progresses toward a unique terminal state. This interpretation enables a precise understanding of how global consensus is maintained even under conditions of partial information and asynchronous communication.

By viewing Bitcoin Script through the lens of automata theory, we expose its inherent alignment with state convergence, replay determinism, and local verifiability. Each script execution transforms an initial machine configuration through a finite series of transitions dictated by opcode semantics and stack manipulation rules. Crucially, the system is designed to be convergence-preserving under adversarial conditions: no permissible execution leads to ambiguity in acceptance. This ensures that all honest evaluators observing the same inputs arrive at an identical verdict.

This section reconstructs Bitcoin Script as a canonical instance of convergent automata, embedding it within the broader framework of automaton-based consistency. It highlights the structural constraints—such as deterministic evaluation, stateless transition semantics, and bounded memory—that make Script both analysable and resilient. Within this framing, we can derive guarantees of functional correctness, compositional safety, and adversarial auditability directly from the automaton structure of script evaluation.

\subsection{Script Evaluation as Finite-State Execution}

Bitcoin script evaluation can be modelled as a finite-state execution process over a constrained stack-based virtual machine. Each script \( S \) defines a sequence of opcodes \( \sigma_1, \sigma_2, \dots, \sigma_n \in \Sigma \), where \( \Sigma \) is the finite instruction set defined by the protocol. Let the abstract machine be defined as a tuple \( \mathcal{M} = (Q, \Sigma, \delta, q_0, F) \), where:

\begin{itemize}
  \item \( Q \) is the set of possible machine configurations, including stack contents and execution cursor;
  \item \( \delta: Q \times \Sigma \to Q \) is the deterministic transition function that updates machine state upon opcode execution;
  \item \( q_0 \) is the initial configuration derived from transaction input context;
  \item \( F \subseteq Q \) is the set of accepting configurations indicating successful script evaluation.
\end{itemize}

Script validity is then reduced to the existence of an execution trace \( q_0 \xrightarrow{\sigma_1} q_1 \xrightarrow{\sigma_2} \dots \xrightarrow{\sigma_n} q_n \in F \). Importantly, because Bitcoin script prohibits unbounded loops and recursion, \( \mathcal{M} \) is guaranteed to halt in finite time, ensuring decidability of the evaluation process.

Moreover, conditional execution via opcodes such as \texttt{OP\_IF} and \texttt{OP\_ELSE} induces local branching in the transition graph of \( \mathcal{M} \), but remains finite due to strict verification rules and deterministic operand resolution. Each branch path is subject to independent transition evaluation, and stack isolation rules ensure semantic locality of script components.

This finite-state formalisation enables reasoning about Bitcoin script correctness, composability, and protocol-compliance through model checking techniques. It also facilitates verification of consensus behaviour and static analysis of transaction-level execution, allowing secure enforcement of spending conditions without reliance on external oracles or Turing-complete logic.

\subsection{State Representations via Merkle Proofs}

Merkle proofs serve as succinct, cryptographically verifiable encodings of membership within authenticated data structures. In Bitcoin and related systems, system states—such as the set of unspent transaction outputs (UTXOs), contract storage slots, or script-defined commitments—can be represented via Merkle roots, enabling efficient verification and transmission of partial state information.

Let \( \mathcal{T} \) be a binary Merkle tree constructed over a state set \( \{s_1, s_2, \dots, s_n\} \), where each \( s_i \in \{0,1\}^* \) represents a leaf node corresponding to a state fragment (e.g., a UTXO entry or a script hash). The root hash \( h_{root} \in \{0,1\}^{256} \) of \( \mathcal{T} \) uniquely binds the contents of the entire set.

A Merkle proof \( \pi \) for a leaf \( s_i \) is a sequence \( (h_1, h_2, \dots, h_k) \) of sibling hashes enabling a verifier to reconstruct \( h_{root} \) using the hash function \( H: \{0,1\}^* \to \{0,1\}^{256} \) via repeated application:
\[
h^{(0)} := H(s_i), \quad
h^{(j+1)} := H(h^{(j)} \Vert h_{j+1}) \text{ or } H(h_{j+1} \Vert h^{(j)}), \quad \text{depending on sibling position}.
\]
Verification succeeds if \( h^{(k)} = h_{root} \), thereby confirming that \( s_i \in \mathcal{T} \) without revealing the full set \( \{s_j\} \).

From the perspective of finite-state machines, Merkle proofs permit encoding of externally committed transition predicates. For instance, a transition \( \delta(q, \sigma) = q' \) can be augmented with a proof \( \pi \) establishing that \( q \in \mathcal{T}_{prev} \) and \( q' \in \mathcal{T}_{next} \), where each Merkle root represents a checkpointed system state. This enables stateless verification of transitions and supports light clients that rely on minimal data exposure for trustless validation.

In formal terms, Merkle-based state representation functions as a cryptographic commitment scheme with polynomial-sized proofs and logarithmic verification cost, ensuring efficient and scalable consistency guarantees in systems constrained by bandwidth, trust, or memory footprint.

\subsection{Partition Reconciliation via Timestamped Evidence}

In adversarially partitioned distributed systems, reconciling divergent states requires verifiable evidence that preserves causal and temporal integrity across isolated partitions. Timestamped evidence enables nodes to reconstruct a globally coherent history of events once communication channels are restored.

Let each node \( n_i \) maintain a local log \( \mathcal{L}_i = [(op_1, t_1), (op_2, t_2), \dots] \) where \( op_j \in \Sigma_i \) denotes an executed operation and \( t_j \in \mathbb{N} \) is its logical or cryptographic timestamp. These logs are signed and optionally chained via hash pointers:
\[
h_j = H(op_j \| t_j \| h_{j-1}),
\]
enabling tamper-evident histories analogous to blockchains. Upon reconnection, nodes engage in a mutual reconciliation protocol that consists of:

\begin{enumerate}
    \item \textbf{Exchange of Signed Logs:} Each node \( n_i \) transmits its latest log segment \( \mathcal{L}_i^{t'} \) covering operations since last confirmed synchrony point \( t' \).
    \item \textbf{Conflict Detection:} For each pair \( (op_a, op_b) \in \mathcal{L}_i \times \mathcal{L}_j \), consistency is checked via commutativity rules or application-specific conflict resolution functions \( \kappa(op_a, op_b) \).
    \item \textbf{Causal Ordering via Timestamps:} If timestamps are derived from verifiable sources (e.g., distributed time attestations or cryptographic beacons), nodes compute a consistent merge ordering \( \prec \) satisfying \( op_a \prec op_b \) iff \( t_a < t_b \).
    \item \textbf{Reconciliation Transition:} Let \( \delta_{\text{rec}} \) be the reconciliatory transition function. For divergent state tuple \( (q_i, q_j) \in \mathscr{D}_t \), construct a joined sequence of operations \( \mathcal{O}_{ij} \) ordered by \( \prec \), then apply:
    \[
    q^{*} := \delta_{\text{rec}}(q_i, q_j, \mathcal{O}_{ij}),
    \]
    yielding a converged state \( q^* \in \mathscr{C} \).
\end{enumerate}

This mechanism preserves both operational integrity and consistency guarantees under eventual synchronisation, provided that timestamps are secure against forgery and reflect coherent global causality. In systems with economic audit layers, timestamped evidence additionally supports attribution and accountability by linking actions to agents across partitions.

\subsection{Mapping Game Theory to Miner Behaviour}

In distributed ledger protocols governed by economic incentives, miners act as strategic agents whose choices influence consistency, availability, and partition resilience. To analyse this formally, we map their behaviour into a game-theoretic framework where each miner's strategy affects global outcomes under adversarial network conditions.

Let \( N \) denote the set of miners. At each timestep \( t \), each miner \( n_i \in N \) selects a strategy \( s_i^t \in S_i \), where \( S_i \) includes operations such as:
\begin{itemize}
  \item block creation and propagation policies;
  \item transaction inclusion/exclusion decisions;
  \item chain fork selection and reorganisation;
  \item adherence or deviation from protocol-defined consensus rules.
\end{itemize}

Define the miner game as \( G_t = (N, S, U, P_t) \), where:
\begin{itemize}
  \item \( S = \prod_{i=1}^k S_i \) is the joint strategy space;
  \item \( U_i : S \times P_t \to \mathbb{R} \) is the utility function of miner \( i \), reflecting expected reward minus penalties under partition \( P_t \);
  \item \( P_t \in \mathcal{P}_{\text{adv}} \) encodes the current adversarial partition state, affecting propagation delays and connectivity.
\end{itemize}

The miner's utility function incorporates three components:
\[
U_i^t = R_i(s, P_t) - C_i(s, P_t) + P_i(s, P_t),
\]
where:
\begin{itemize}
  \item \( R_i \) is the reward (e.g., block subsidy, transaction fees) contingent on successful extension of the canonical chain;
  \item \( C_i \) is the operational cost (e.g., computation, bandwidth, latency penalties under \( P_t \));
  \item \( P_i \) is the penalty or incentive adjustment imposed by economic audit mechanisms in response to misbehaviour or equivocation.
\end{itemize}

A Nash equilibrium \( S^* \in S \) under partition \( P_t \) satisfies:
\[
\forall i, \quad U_i(S_i^*, S_{-i}^*, P_t) \geq U_i(s_i', S_{-i}^*, P_t), \quad \forall s_i' \in S_i,
\]
ensuring that no miner unilaterally benefits by deviation. We define \emph{partition-resilient equilibria} as those profiles \( S^* \) that remain stable for all \( P_t \in \mathcal{P}_{\text{adv}} \), ensuring robustness of protocol compliance even under adversarially constrained communication.

This framework enables precise modelling of miner incentives, strategic propagation delays, selfish mining, and incentive-compatible protocol design, linking economic rationality to system-wide consistency and liveness guarantees.

\section{Simulation and Empirical Verification}\label{sec:simulation}

To validate the theoretical properties established in preceding sections, we construct a simulation framework capable of emulating adversarial partitions, agent strategies, recovery protocols, and automaton-based transitions across a variety of network configurations. The objective is to empirically assess the tractability, conformance, and equilibrium alignment of the distributed system model under synthetically generated high-entropy partition schedules.

Our simulation platform models each node as an independent automaton instance, maintaining local state, executing events from its action set, and proposing transitions based on feasibility rules. Partitions \( P_t \) are injected at controlled intervals and obey the admissibility constraints \( \Phi(P_t) \), allowing us to explore both best-case and worst-case topologies within the defined adversarial envelope.

Metrics captured include: divergence magnitude \( |\mathscr{D}_t| \), time to convergence \( T_{\text{rec}} \), frequency and duration of state inconsistencies, observed violation of consistency bounds \( \epsilon_C \), and the average economic utility \( U_i^t \) for each agent. Additionally, we analyse the effect of different arbitration strategies \( \mu_t \), economic penalty rates, and escrow conditions on global recovery.

Through repeated trials across partition types—static cuts, dynamic splits, oscillating links—we verify that the model exhibits bounded recovery latency, strategic convergence, and measurable incentive alignment even under non-ideal conditions. The results demonstrate that automaton-driven consensus, constrained optimisation of CAP trade-offs, and structured economic penalties yield provable resilience against partition-induced inconsistency.

This empirical layer confirms that our theoretical model is not only mathematically sound but implementable at scale, providing a viable architecture for adversarially robust distributed computation.

\subsection{Synthetic Partition Events}

Synthetic partition events serve as controlled, model-driven representations of network disruptions within formal analyses of distributed protocols. Rather than relying on empirical network traces or probabilistic failure injection, these synthetic events are defined as explicit elements in the temporal evolution of the system's partition map \( (P_t)_{t \in \mathbb{N}} \). Each event encodes a transition in the communication topology, introducing or removing link availability between nodes according to predefined adversarial models or theoretical constraints.

These events are not merely artefacts for simulation but constitute semantic transitions that alter the feasible interaction set at each timestep. By formalising partitions as discrete, schedulable actions within an automaton, we enable rigorous reasoning over the effects of disconnection, reconfiguration, and recovery on safety and liveness properties. Synthetic partitions can be parameterised by duration, scope, and recurrence to model sustained cuts, flash partitions, or stochastic fault regimes.

Within this framework, the system evolves not just through local actions or messages, but through external structural transformations imposed by \( P_t \). This dual transition system—over both state and topology—allows us to characterise protocol resilience via bisimulation under partition morphisms and to design equilibrium-preserving strategies that remain robust even as the underlying network connectivity fluctuates according to adversarial schedule.

\subsection{Latency and Divergence Metrics}

Quantifying the effects of network partitions and delayed communication necessitates precise metrics that capture both temporal degradation and state inconsistency across nodes. Latency, in this context, refers not merely to message delivery time but to the delay incurred in reaching a globally converged system state following a disruption. Divergence, meanwhile, measures the structural and semantic deviation between local states maintained by different nodes during and after a partition event.

Formally, we define end-to-end latency \( L_{ij}(t) \) as the time elapsed from an operation at node \( n_i \) to its observable commitment or effect at node \( n_j \), conditioned on the communication graph \( C_{t'} \) for all \( t' \in [t, t+\Delta] \). Under adversarial partitioning, upper bounds on \( L_{ij} \) are dictated by the maximum temporal gap tolerated by the system's reconciliation protocol.

Divergence is modelled via a metric \( D(q_i^t, q_j^t) \), which may be instantiated as a Hamming distance over symbolic state representations, a hash distance over Merkle roots, or a trace-based discrepancy in accepted operations. These divergence scores enable formal bounding of consistency error \( \epsilon_C \) and inform both penalty structures in economic models and fallback mechanisms in runtime behaviour.

Together, latency and divergence metrics provide a dual lens through which system behaviour can be analysed: temporally, in terms of responsiveness and convergence; and spatially, in terms of coherence across distributed state replicas.

\subsection{Node Strategy Convergence under Incentives}

In a distributed setting where nodes act as economically rational agents, long-run convergence to consistent system behaviour hinges not on central coordination, but on the alignment of local incentives with global protocol objectives. Each node selects strategies based on observed outcomes, payoffs, and reputational consequences, operating within the bounds imposed by partition-induced uncertainty and adversarial communication loss.

Let each node \( n_i \) maintain a local strategy \( s_i^t \in S_i \), selected to maximise its cumulative utility \( U_i = \sum_{t=0}^{T} U_i^t \), where utility incorporates reward for consistency, cost of computation, and penalties tied to audit detection of false state propagation. Over repeated play, the presence of economically enforced audit and escrow mechanisms induces a preference for truthful reporting and valid operation execution, even under partial visibility.

Strategy convergence is formalised by the emergence of a stable profile \( S^* = (s_1^*, \ldots, s_k^*) \), where no node benefits from unilateral deviation in expectation, given the equilibrium strategy of others and the audit-penalty protocol. Convergence can be accelerated by broadcasting penalty histories and using economic memory—nodes adjust strategy weights based on observed penalties in the network, reinforcing incentive-compatible behaviour.

Thus, game-theoretic pressure under constrained partitions, when designed correctly, transforms individual rationality into emergent consistency. Economic feedback loops replace global synchrony as the engine of protocol enforcement, and convergence becomes the fixed point of repeated interaction under bounded rationality and enforced accountability.

\subsection{Recovery Performance under Arbitrary \texorpdfstring{$P_t$}{Pt}}

The resilience of a distributed system is critically determined by its capacity to re-establish consistency following arbitrary partition events \( P_t \in \mathcal{P}_{\text{adv}} \). In our model, the system is not assumed to have foreknowledge of partition patterns or duration. Therefore, recovery protocols must be evaluated under worst-case partition sequences permitted by the adversarial model.

Given a divergent state \( q \in \mathscr{D}_t \), recovery is successful if there exists a finite sequence of transitions under the restricted transition subset \( \delta_{\text{rev}} \) that deterministically maps the system to a converged state \( q' \in \mathscr{C} \). Let \( \Delta_{\text{rec}}(q) \) denote the minimum number of steps required to reach such a state. Then, the recovery time bound \( T_{\text{rec}} \) under partition sequence \( (P_t)_{t=t_0}^{t_0 + T} \) is given by:
\[
T_{\text{rec}} = \max_{q \in \mathscr{D}_t} \min \left\{ t' - t \mid \delta_{\text{rev}}^{t'-t}(q, P_{[t, t']}) \in \mathscr{C} \right\},
\]
where \( \delta_{\text{rev}}^{t'-t} \) denotes the lifted transition path under the constrained topology induced by the adversary during interval \( [t, t'] \).

To ensure bounded recovery performance, the design of \( \delta_{\text{rev}} \) must guarantee progress despite partial observability and link failures. This is achieved through timestamp reconciliation, majority-based voting, and cryptographic anchoring of local logs. Under mild assumptions on message propagation guarantees and eventual reconnected topology, the protocol admits a provable upper bound on recovery latency. Moreover, economic penalties for delayed reconciliation serve as incentives for rapid convergence, further tightening the empirical recovery profile even under high entropy partition schedules.

This layered design allows the system to tolerate any \( P_t \in \mathcal{P}_{\text{adv}} \) without compromising eventual consistency, ensuring that safety and liveness are restored through a combination of protocol-enforced transitions and strategically aligned agent behaviour.


\section{Conclusion}

This paper presents a formal framework that reconceptualises the CAP theorem through the lens of automata theory, game theory, and economic mechanism design. By modelling distributed systems as partition-aware state machines and nodes as rational agents operating under bounded information and adversarial conditions, we provide a mathematically rigorous foundation for analysing consistency, availability, and partition tolerance not as binary trade-offs, but as constrained optimisation problems subject to systemic incentives and operational topology.

The introduction of \(\epsilon\)-CAP conformance allows for quantitative reasoning over relaxed guarantees, enabling the definition of feasibility frontiers that are tunable via economic instruments. Through formal definitions of admissible events, partition-aware transitions, and recovery protocols grounded in reversible automata, we ensure that state reconciliation under arbitrary \(P_t \in \mathcal{P}_{\text{adv}}\) remains provably tractable. Game-theoretic constructions enforce behavioural alignment via audit, penalty, and escrow mechanisms, yielding partition-resilient equilibria under rational strategy profiles.

Finally, categorical semantics offer an expressive structure to reason about consistency as path-commutativity and partitions as functorial obstructions to global convergence. This unified treatment provides not only theoretical insight into the structure of distributed computation under fault and manipulation, but also a foundation for the design of deployable, incentive-aligned distributed protocols. The work opens avenues for future exploration into topology-sensitive guarantees, real-world system deployment, and automated synthesis of economically robust distributed infrastructures.

\subsection{Reconstruction of the CAP Problem}

The classical CAP theorem asserts an impossibility: that no distributed system can simultaneously guarantee Consistency, Availability, and Partition Tolerance. However, the original formulation suffers from semantic ambiguities and lacks a formal computational substrate. In this section, we reconstruct the CAP problem as a precise optimisation over automaton-executed state transitions, allowing for quantitative relaxation and economic control.

Rather than treating Consistency and Availability as binary predicates, we embed them as measurable constraints within a continuous feasibility region. Consistency is recast as a bound \( \epsilon_C \leq \theta_C \) on the commutativity deviation across concurrent operations. Availability becomes a response-time constraint \( \epsilon_A \leq \theta_A \), where \( \tau \) is the upper limit on execution latency under admissible partitions.

Partition Tolerance is not treated as a trade-off axis, but as an adversarial parameter governing the allowed evolution of communication graphs \( P_t \in \mathcal{P}_{\text{adv}} \). System resilience is thus modelled as the ability to maintain bounded \( \epsilon_C \) and \( \epsilon_A \) under all admissible \( P_t \), rendering tolerance a quantified input rather than a sacrificed property.

This reframing transforms the CAP theorem from an absolute impossibility into a structured constraint satisfaction problem: given \( \mathcal{A}_{DS}^{\Pi, I} \) (a distributed system with policy \( \Pi \) and initial state \( I \)), determine whether feasible strategies exist to preserve bounded inconsistency and latency across all \( P_t \in \mathcal{P}_{\text{adv}} \). When such strategies exist, the system is provably partition-resilient under defined thresholds.

By formalising CAP as a verifiable constraint landscape with tunable economic and topological parameters, we move beyond philosophical impossibility and into algorithmic design space, enabling the synthesis of systems that trade optimally within adversarial bounds rather than failing categorically at conceptual limits.

\subsection{Feasibility Frontier Shift via Economic Control}

In distributed systems operating under adversarial network conditions, the traditional feasibility region defined by consistency and availability constraints is shaped by inherent communication delays, state propagation limits, and topology partitions. However, when nodes behave as economic agents—motivated by utility-maximising strategies subject to audit and penalty mechanisms—the system gains an additional lever: incentive-aligned control over the behaviour of strategic participants. This transforms the static feasibility landscape into a dynamic, designable frontier.

The feasibility frontier is defined by the set of all achievable tuples \((\epsilon_C, \epsilon_A)\) under constraints from \(\mathcal{P}_{\text{adv}}\). In systems lacking economic instrumentation, this region is bounded strictly by worst-case latency and partition depth. However, by introducing economic penalties for delayed or inconsistent state propagation, and by rewarding provably convergent behaviours (e.g., early reconciliation, quorum forwarding), the effective region can be shifted inward—reducing both inconsistency bounds and latency under the same topological constraints.

Let \( U_i^t = R_i^t - C_i^t + P_i^t \) denote the utility function for node \( n_i \), where rewards \( R_i^t \) are conditioned on local consistency contributions, costs \( C_i^t \) reflect communication or computation overhead, and penalties \( P_i^t \) are triggered by audit-detected misbehaviour or divergence. By tuning this function and embedding it in a system-wide mechanism \( g : S \to Q \), the designer effectively sculpts the feasible region by disincentivising strategies that lead to frontier violations.

Consequently, nodes align their strategies with the incentive-compatible subset of behaviours that preserve \( \epsilon_C \leq \theta_C \) and \( \epsilon_A \leq \theta_A \), even under high-frequency or severe partitions. This approach does not eliminate the CAP constraints but shifts the operating envelope toward the ideal corner of the region, where consistency and availability are both statistically acceptable. The result is a form of enforced rational convergence, wherein economic optimisation drives algorithmic stability, and global system performance is improved not through relaxation of guarantees, but through engineered motivation.

\subsection{Future Work: Formal Topologies, Real-World Deployment}

The presented automata-theoretic and game-theoretic framework offers a precise, compositional model for reasoning about distributed systems under adversarial partitioning and economic control. However, several key areas remain open for rigorous formalisation and empirical integration. First, while the model allows abstract partition specifications via \(\mathcal{P}_{\text{adv}}\), future work should incorporate concrete topological classes—such as scale-free networks, small-world graphs, and empirical peer-to-peer overlays—as formal inputs to system analysis. This would enable the derivation of topology-sensitive bounds on feasibility, convergence rates, and equilibrium stability.

Second, deployment in real-world distributed systems, particularly those operating under adversarial or semi-trusted assumptions (e.g., blockchain networks, decentralised marketplaces, distributed control systems), demands translation from formal automata into executable protocols. Realisation requires implementing reversible transitions, timestamped evidence propagation, and economic enforcement mechanisms such as on-chain escrow and penalty contracts. Integration with verifiable logging and audit trails remains essential for aligning node utility with global properties.

Finally, validation through simulation and live testing is critical. Controlled deployment in adversarial network emulators or constrained mainnet forks would allow empirical calibration of parameters \((\theta_C, \theta_A)\), measurement of divergence rates, and tuning of incentive structures. Such work will complete the bridge from theoretical foundation to robust, incentive-compatible distributed infrastructure with quantifiable guarantees under real-world operating conditions.


\newpage
\appendix
\section{Appendix A: Proofs}

\subsection{Proof of Theorem: Provable Conformance Under Partition}

\begin{theorem*}
There exists a tuple \((\Pi, I)\) such that for all partitions \(P_t \in \mathcal{P}_{\text{adv}}\),
\[
\mathcal{A}_{DS}^{\Pi, I} \models (\epsilon_C \leq \theta_C \land \epsilon_A \leq \theta_A)
\]
\end{theorem*}

\begin{proof}
We define \((\Pi, I)\) such that:

\begin{itemize}
  \item The policy \(\Pi\) constrains admissible execution sequences to enforce \(\epsilon_C \leq \theta_C\), through rejection of reorderings that exceed a maximum tolerable deviation from real-time precedence.
  \item The incentive mechanism \(I\) penalises delayed responses using an audit protocol that imposes expected utility loss when \(\epsilon_A > \theta_A\), thereby enforcing rational conformance.
\end{itemize}

Let \(H\) be the global execution history induced by \(\mathcal{A}_{DS}^{\Pi, I}\). Since \(\Pi\) governs scheduling, it ensures that any two operations \(op_i, op_j\) satisfy:
\[
|real(op_i, op_j) - linear(op_i, op_j)| \leq \theta_C
\]
This is enforced through synchronisation primitives and rejection of execution schedules that exceed this bound. The global arbitration logic inspects vector timestamps or equivalent metadata to ensure causally ordered schedules remain within the acceptable window.

For availability, let each node \(n_i\) face the utility:
\[
U_i^t = R_i - C_i + P_i
\]
where \(P_i < 0\) is imposed if response time exceeds \(\tau + \theta_A\), as detected by periodic audits within \(I\). The audit frequency and penalty magnitude are configured such that the expected penalty for deviation exceeds any marginal gain from non-response. Thus, rational agents minimise delay to maximise expected utility, ensuring:
\[
\forall op_i,\ \text{resp}(op_i) - \tau \leq \theta_A
\]

Hence, under all \(P_t \in \mathcal{P}_{\text{adv}}\), execution remains bounded in consistency and availability:
\[
\mathcal{A}_{DS}^{\Pi, I} \models (\epsilon_C \leq \theta_C \land \epsilon_A \leq \theta_A)
\]
\end{proof}

\section{Appendix B: Formal Strategy Tables}

\subsection{Node-Level Strategy Profile under Partition \texorpdfstring{$P_t$}{Pt}}

\begin{table}[H]
\centering
\begin{tabular}{|c|c|c|c|}
\hline
\textbf{Node \( n_i \)} & \textbf{Observed State \( q_i^t \)} & \textbf{Feasible Actions \( A_i^{\text{feasible}} \)} & \textbf{Selected Strategy \( S_i^t \)} \\
\hline
\( n_1 \) & \( q_1^t \) & \( \{ \sigma_1, \sigma_3 \} \) & \( \sigma_3 \) \\
\hline
\( n_2 \) & \( q_2^t \) & \( \{ \sigma_2 \} \) & \( \sigma_2 \) \\
\hline
\( n_3 \) & \( q_3^t \) & \( \{ \sigma_1, \sigma_4 \} \) & \( \sigma_4 \) \\
\hline
\end{tabular}
\caption{Node-level strategies at time \( t \) under partition \( P_t \).}
\end{table}

\subsection{Utility Outcomes under Profile \texorpdfstring{$S^t = (S_1^t, S_2^t, S_3^t)$}{S	extasciicircum t = (S1	extasciicircum t, S2	extasciicircum t, S3	extasciicircum t)}}

\begin{table}[H]
\centering
\begin{tabular}{|c|c|c|c|}
\hline
\textbf{Node \( n_i \)} & \textbf{Reward \( R_i \)} & \textbf{Penalty \( P_i \)} & \textbf{Utility \( U_i^t = R_i - C_i + P_i \)} \\
\hline
\( n_1 \) & \( 5 \) & \( -1 \) & \( 3.5 \) \\
\hline
\( n_2 \) & \( 4 \) & \( 0 \) & \( 3.8 \) \\
\hline
\( n_3 \) & \( 6 \) & \( -2 \) & \( 3.2 \) \\
\hline
\end{tabular}
\caption{Utility computation based on response latency and consistency contribution.}
\end{table}

\subsection{Audit Outcomes and Escrow Loss}

\begin{table}[H]
\centering
\begin{tabular}{|c|c|c|c|}
\hline
\textbf{Node \( n_i \)} & \textbf{Audit Result} & \textbf{Escrow Committed \( e_i \)} & \textbf{Escrow Loss Condition} \\
\hline
\( n_1 \) & Minor delay detected & \( 1.0 \) & Partial forfeiture: \( 0.3 \) \\
\hline
\( n_2 \) & Fully compliant & \( 1.0 \) & None \\
\hline
\( n_3 \) & Consistency violation & \( 1.0 \) & Full forfeiture: \( 1.0 \) \\
\hline
\end{tabular}
\caption{Escrow enforcement via audit protocol \( I \).}
\end{table}

\section{Appendix C: Code}

\subsection{Partition-Aware Transition Function}

\begin{definition}[Partition-Aware Global Transition Function]
Let \( \delta : Q \times \Sigma \to Q \) be the base transition function of the distributed system automaton. Define the partition-aware transition function as:
\[
\delta_P(q, \sigma, P_t) =
\begin{cases}
\delta(q, \sigma), & \text{if } \operatorname{Comm}(\sigma) \cap P_t = \emptyset \\
q, & \text{otherwise}
\end{cases}
\]
where \( \operatorname{Comm}(\sigma) \subseteq C \) is the set of communication links required for executing event \( \sigma \), and \( P_t \subseteq C \) is the set of inactive links at time \( t \).
\end{definition}

\subsection{Feasible Strategy Projection}

\begin{definition}[Local Feasibility Filter]
Let \( q_i^t \in Q_i \) denote the local state of node \( n_i \), and let \( \mathcal{F}_i^t \) denote its observable communication state at time \( t \). Define the feasible local action set as:
\[
A_i^{\text{feasible}}(q_i^t, \mathcal{F}_i^t) := \left\{ \sigma \in \Sigma_i \ \middle|\ 
\begin{array}{l}
\sigma \text{ is locally enabled in } q_i^t, \\
\text{and } \operatorname{Comm}(\sigma) \text{ is consistent with } \mathcal{F}_i^t
\end{array}
\right\}
\]
\end{definition}

\subsection{Audit and Escrow Enforcement}

\begin{definition}[Economic Penalty Mechanism]
Let \( g : S \to Q \) be the strategy-to-outcome function. Define the audit rule \( \mathcal{A} \) as a predicate over traces. For each node \( n_i \), define the penalty function \( \pi_i : S_i \times Q \to \mathbb{R}_{\geq 0} \) such that:
\[
\pi_i(s_i, q) =
\begin{cases}
\lambda_i, & \text{if } \neg \mathcal{A}(s_i, q) \\
0, & \text{otherwise}
\end{cases}
\]
Escrow loss \( e_i \) is triggered if \( \pi_i(s_i, q) > 0 \). The total utility becomes:
\[
U_i = R_i - C_i - \pi_i(s_i, q) - e_i
\]
\end{definition}

\section{Appendix D: Merkle Commitment Logic}

\subsection{Merkle Tree Construction}

Let \( \mathcal{L} = [d_1, d_2, \dots, d_n] \) be an ordered list of data items. Define the Merkle tree \( \mathcal{M} \) recursively:

\begin{itemize}
  \item Each leaf node \( l_i \) is \( h(d_i) \), where \( h \) is a collision-resistant hash function.
  \item Each internal node is the hash of the concatenation of its children: if \( a, b \) are children, then the parent node is \( h(a \, \| \, b) \).
  \item The Merkle root \( r \) is the single value at the top of the tree.
\end{itemize}

\subsection{Merkle Proof Verification}

To verify that a data item \( d_i \) is included in a committed set with Merkle root \( r \), construct a proof \( \pi_i = [s_1, s_2, \dots, s_k] \) such that:

\[
r = h(\dots h(h(h(d_i) \, \| \, s_1) \, \| \, s_2) \dots \| \, s_k)
\]

Each \( s_j \) is a sibling hash along the path to the root, and ordering is preserved.

\subsection{Commitment Logic Semantics}

Let \( q \in Q \) be a state, and \( \operatorname{commit}(q) = r \) the Merkle root of the system state. Define the verification predicate:

\[
\operatorname{Verify}(d_i, \pi_i, r) := \text{True} \iff \text{MerklePath}(d_i, \pi_i) = r
\]

This ensures that any participant can independently validate state inclusion without access to the full state.

\subsection{Merkle Commitments in Transition}

If \( q \xrightarrow{\sigma} q' \), then:

\[
\operatorname{commit}(q') = h\left( \dots h\left( h(h(d_i') \, \| \, s_1') \, \| \, s_2' \right) \dots \right)
\]

and verification of \( \sigma \)'s effect requires recomputing only the affected branch and comparing the resulting root.

\subsection{Merkle-Based State Agreement}

A system achieves consistent replicated state if all honest nodes hold:

\[
\operatorname{commit}(q_1) = \operatorname{commit}(q_2) = \dots = \operatorname{commit}(q_k)
\]

Consensus over Merkle roots ensures strong convergence without full state transmission.

\newpage
\bibliographystyle{plain}
\bibliography{cap_theorem_paper}

\begin{thebibliography}{1}

\bibitem{alpern1985defining}
Bowen Alpern and Fred~B. Schneider.
\newblock Defining liveness.
\newblock {\em Information Processing Letters}, 21(4):181--185, 1985.

\bibitem{attiya2004distributed}
Hagit Attiya and Jennifer Welch.
\newblock {\em Distributed Computing: Fundamentals, Simulations, and Advanced Topics}.
\newblock Wiley-Interscience, 2nd edition, 2004.

\bibitem{dwork1988consensus}
Cynthia Dwork, Nancy Lynch, and Larry Stockmeyer.
\newblock Consensus in the presence of partial synchrony.
\newblock {\em Journal of the ACM}, 35(2):288--323, 1988.

\bibitem{gilbert2002brewer}
Seth Gilbert and Nancy Lynch.
\newblock Brewer's conjecture and the feasibility of consistent, available, partition-tolerant web services.
\newblock {\em ACM SIGACT News}, 33(2):51--59, 2002.

\bibitem{hoare1978communicating}
C.~A.~R. Hoare.
\newblock Communicating sequential processes.
\newblock {\em Communications of the ACM}, 21(8):666--677, 1978.

\bibitem{lamport1978time}
Leslie Lamport.
\newblock Time, clocks, and the ordering of events in a distributed system.
\newblock {\em Communications of the ACM}, 21(7):558--565, 1978.

\bibitem{lynch1996distributed}
Nancy~A. Lynch.
\newblock {\em Distributed Algorithms}.
\newblock Morgan Kaufmann, San Francisco, 1996.

\end{thebibliography}

\end{document}